\documentclass[aps,pra,reprint,superscriptaddress,nofootinbib,longbibliography,floatfix]{revtex4-2}
\usepackage{amsmath,amssymb,graphicx,mathptmx}
\usepackage[hidelinks]{hyperref}
\usepackage{needspace}
\begin{document}
\raggedbottom
\title{Optimizing interference zeros for finite-pulse transmission in waveguide quantum electrodynamics}
\author{J. Zhou}
\affiliation{College of Physics and Materials Science, Tianjin Normal University, Tianjin 300387, China}
\author{Y. M. Zhang}
\affiliation{College of Physics and Materials Science, Tianjin Normal University, Tianjin 300387, China}
\author{X. Z. Zhang}
\email{zhangxz@tjnu.edu.cn}
\affiliation{College of Physics and Materials Science, Tianjin Normal University, Tianjin 300387, China}
\affiliation{Interdisciplinary Center, Tianjin Normal University, Tianjin 300387, China}

\begin{abstract}
A photon pulse occupies a range of frequencies, so suppressing reflection at its carrier need not minimize the error of the transmitted pulse. We study two separated emitters coupled to a common waveguide at several contact points. Interference between the contact paths can make the coupling that drives each emitter quadratic in frequency detuning. With matched propagation and emitter resonances centered on the carrier, the reflected fields cancel once more. Their combined reflection probability is then sixth order at zero loss. Separating each local double zero into two simple zeros can nevertheless improve transmission of a finite pulse. We minimize the leading narrowband error within this common static adjustment, counting reflection, absorption, and distortion of the transmitted mode. For a Gaussian pulse without loss, the minimum is two-fifths of the error for coincident zeros. Independent emitter loss favors a smaller separation because photons entering separate reservoirs do not share the cancellation of the reflected fields. The optimum follows from the pulse spectrum and the relative weight of these two processes. Uniform and Su--Schrieffer--Heeger waveguides realize the mechanism. In the latter, coordinated contact and resonance adjustments also preserve the full ratio of complex transmission amplitudes for opposite incidence directions. Finite-band calculations determine how the benefit changes with bandwidth and contact errors.
\end{abstract}
\maketitle

\section{Introduction}
A localized emitter can strongly alter the transmission of a single photon in a one-dimensional waveguide \cite{Shen2005,Fan2010}. Experiments with superconducting artificial atoms have realized this interaction \cite{Astafiev2010}, providing control over propagating photons and collective emitter dynamics \cite{Roy2017,Sheremet2023}. For two separated emitters, the photon can scatter from both sites before leaving the waveguide. The phase accumulated between the emitters determines how their radiated fields interfere, producing transmission resonances and collective radiation \cite{Tsoi2008,Lalumiere2013}. Photon-mediated interactions have also been observed between distant superconducting qubits \cite{vanLoo2013}.

An emitter coupled to the waveguide at several points offers a second source of interference. A photon can excite that emitter through different contacts, and the corresponding amplitudes add before the excitation probability is formed. Their relative phases and strengths therefore change both the radiative decay rate and the emitter frequency shift \cite{Kockum2014,Kockum2018}, as realized with superconducting artificial atoms \cite{Kannan2020}. Propagation between contacts can also produce delayed feedback when the travel time is comparable to the emitter lifetime \cite{Guo2017}. The same interference can distinguish the two propagation directions \cite{Lodahl2017,Soro2022}. These effects allow the contacts to shape the response across a pulse spectrum, as well as at one selected frequency.

Quantum state transfer requires control of the photon's temporal mode \cite{Cirac1997}, and microwave techniques can generate single photons with prescribed pulse shapes \cite{Pechal2014}. Multipoint emitters also support directional emission over a broad band \cite{Wang2024} and state transfer through static coupling profiles \cite{Diekmann2026}. For transmission through a scatterer, high transmitted power is not sufficient: a frequency-dependent phase can distort the outgoing pulse. Frequencies at which reflection vanishes provide a route to impedance matching \cite{Sweeney2020}. Higher-order output zeros have been related to the residual scattering of Gaussian wave packets \cite{Li2026}, and distributing scattering zeros across a band can improve finite-band routing \cite{Faul2025}. These results motivate a specific question. For a given incident pulse, should the reflection zeros coincide at the carrier or be separated across the occupied spectrum? Independent emitter loss can change the answer because its frequency dependence differs from that of coherent reflection.

Here we determine how two emitters should be adjusted to preserve a prescribed pulse incident from one side. At the reference setting, the contact paths make the coupling amplitude of each emitter vanish quadratically at the carrier. When the propagation phase is matched and the mean shifted emitter frequency lies at the carrier, cancellation between the reflected fields adds one further power of detuning at zero loss. The total reflected amplitude is then cubic. We adjust both emitters identically to split each local double zero into two simple zeros, while the collective reflection zero remains at the carrier in the lossless case. This redistributes reflection across the pulse spectrum. Photons lost by different emitters enter separate reservoirs, so their loss probabilities add without the extra cancellation of the reflected field. This difference gives a leading optimal separation determined by moments of the incident spectrum. For a Gaussian pulse without loss, the leading mode error falls to two-fifths of its value for coincident contact zeros. Independent loss shifts the optimum toward a smaller separation. The derivation retains frequency-dependent radiative rates and emitter frequency shifts and applies to the common static contact family defined in Sec.~\ref{sec:general}.

A waveguide with alternating nearest-neighbor couplings provides a concrete realization. This Su--Schrieffer--Heeger (SSH) chain \cite{Su1979} has two sites per unit cell, so an emitter coupled to both sites can be excited through two interfering paths. Earlier work has explored cooperative radiation in dimerized waveguides \cite{Bello2019}, realized such waveguides in superconducting circuits \cite{Kim2021}, and shown how contact geometry shapes two-emitter scattering \cite{Zhu2025}. Exact treatments of two-qubit SSH scattering \cite{Zhou2026} and phase-dependent single-qubit wave-packet scattering \cite{Zhang2026} provide further background. In the geometry considered here, coordinated changes of the contact phases, strengths, and emitter frequencies separate the local zeros while keeping the ratio of opposite-incidence transmission amplitudes fixed at every frequency in the band. At fixed transmission ratio, allowing unequal A/B strengths or different adjustments of the two emitters cannot lower the leading error near the reference contacts. A uniform chain with three contacts per emitter realizes the same pulse optimization. It shows that the mechanism does not depend on the special SSH identities.

The remainder of this paper is organized as follows. Section~\ref{sec:general} develops the local and collective interference mechanisms. Section~\ref{sec:generalpulse} derives the pulse-dependent zero separation and its dependence on loss, and illustrates its realization in a uniform chain. Section~\ref{sec:ssh} gives the SSH realization and the condition for preserving the ratio of directional transmission amplitudes. Sections~\ref{sec:finite} and \ref{sec:calibration} examine finite-bandwidth performance and calibration errors, and Sec.~\ref{sec:conclusion} discusses the results. Appendix~\ref{app:general} derives the scattering and pulse expansions. Appendix~\ref{app:realizations} gives the microscopic waveguide calculations, and Appendix~\ref{app:optimality} treats the allowed contact variations. Appendices~\ref{app:finitepulse} and \ref{app:contacterrors} give the finite-pulse comparisons and the bounds in the presence of contact errors.

\section{Interference in scattering by two emitters}
\label{sec:general}
A photon enters the waveguide from the left and encounters two separated emitters. Each emitter couples to the waveguide at several contact points. The amplitudes for exciting one emitter through these contacts can cancel. The fields reflected by the two emitters can also cancel because they enter the same outgoing waveguide channel. Loss into two independent reservoirs is different: the final reservoir states are orthogonal, so their probabilities add. These different ways of combining amplitudes determine how moving the contact zeros changes reflection and loss.

\subsection{Geometry, coupling amplitudes, and loss}
We set $\hbar=1$ and consider one photon scattering from two emitters in an otherwise lossless waveguide. Each emitter has one excited state. An open energy interval containing the carrier $E_0$ supports one right-moving and one left-moving channel, with finite nonzero group speeds. The contact regions have the same shape and are displaced along the waveguide without overlap. Their coupling strengths and phases are not yet assumed equal. Propagation between the regions contributes the phase $p(E)$, with $|p(E)|=1$, in either direction.

The scattering construction assumes that these two traveling channels account for all coupling between the contact regions. Additional propagating channels, or an appreciable evanescent field connecting the regions, require a larger scattering description. The scalar construction is exact for the lattice realizations used here.

Let $b_{jR}(E)$ be the coupling coefficient with which a right-moving photon drives emitter $j$, and define $b_{jL}(E)$ similarly for a left-moving photon. We normalize the traveling waves to unit flux. The squared amplitudes $u_j=|b_{jR}|^2$ and $g_j=|b_{jL}|^2$ are the radiative rates associated with the two traveling channels. Reflection involves excitation from one direction and emission into the other, so it contains the product $c_j=b_{jL}^*b_{jR}$.

The waveguide also shifts the emitter transition energy by $\Lambda_j(E)$, the real part of its retarded self-energy. For a bare transition energy $\omega_j$, the response denominator is
\begin{equation}
 D_j=E-\omega_j-\Lambda_j(E)+\frac{i}{2}(\gamma+u_j+g_j).
 \label{eq:generalD}
\end{equation}
Each emitter additionally loses population to its own Markov reservoir through the jump operator $\sqrt\gamma\,\sigma_j^-$. The rate $\gamma\ge0$ is the same for both emitters. The waveguide-induced rates and frequency shift retain their full energy dependence. Eliminating the waveguide gives the single-emitter scattering amplitudes
\begin{align}
 t_{jL}&=1-\frac{iu_j}{D_j},&t_{jR}&=1-\frac{ig_j}{D_j},\nonumber\\
 r_{jL}&=-\frac{ic_j}{D_j},&r_{jR}&=-\frac{ic_j^*}{D_j}.
 \label{eq:single}
\end{align}
Here $L$ and $R$ on a scattering amplitude label the side of incidence. A photon incident from the left is initially right-moving and produces the emitter excitation amplitude $b_{jR}/D_j$. Emission into the left-moving channel contributes the conjugate coupling $b_{jL}^*$, giving $r_{jL}=-ib_{jL}^*b_{jR}/D_j$. The transmitted field contains both the incident photon and the forward-emitted field, which explains the unit term in $t_{jL}$. The flux convention gives $b_{jR,L}$ units of the square root of energy, so $u_j$, $g_j$, and $\gamma$ are rates when $\hbar=1$.

A photon can reflect back and forth between the emitters before it leaves. The phase for one round trip is $P(E)=p(E)^2$. For unit incidence from the left, let $f$ be the right-moving field just after emitter 1 and $b$ the left-moving field returning to it. The local scattering relations are
\begin{equation}
 f=t_{1L}+r_{1R}b,\qquad b=Pr_{2L}f.
 \label{eq:roundtripfields}
\end{equation}
Solving them gives $f=t_{1L}/Z$, where $Z=1-Pr_{1R}r_{2L}$. The outgoing fields on the right and left are $pt_{2L}f$ and $r_{1L}+t_{1R}b$, respectively. After removing the phase acquired in an empty waveguide, the result is
\begin{align}
 t_L&=\frac{t_{1L}t_{2L}}{Z},&t_R&=\frac{t_{1R}t_{2R}}{Z},\nonumber\\
 r_L&=r_{1L}+\frac{Pt_{1L}t_{1R}r_{2L}}{Z}.
 \label{eq:cascade}
\end{align}
Equation~(\ref{eq:cascade}) includes every return between the emitters and retains the full propagation phase $p(E)$. We expand this response near the carrier only after the local and repeated-scattering contributions have been combined.

For incidence from side $\alpha$, the transmission and reflection probabilities are $T_\alpha=|t_\alpha|^2$ and $R_\alpha=|r_\alpha|^2$. Let $q_{j\alpha}$ denote the excitation amplitude of emitter $j$ for unit incident flux. The probability lost to the two reservoirs is $A_\alpha=\gamma\sum_j|q_{j\alpha}|^2$. Their orthogonal final states require these probabilities to add, even though the emitter excitations are generated coherently through the same waveguide. The reflected fields instead add before the squared modulus is taken. The self-energy relation in Appendix~\ref{app:general} gives $T_\alpha+R_\alpha+A_\alpha=1$. At zero loss, the two-channel scattering matrix is unitary and $T_L=T_R$, although the transmission phases can differ. Different transmission probabilities require loss in this setting \cite{Chen2022}.

\subsection{Cancellation at a single emitter}
We now choose the same contact strengths and phases for both emitters and write their common amplitudes as $b_R,b_L$. At the carrier $E_0$, the contact paths cancel the right-moving drive. We require the cancellation to persist through first order in the energy detuning $x=E-E_0$. The leading coupling amplitude is then quadratic in $x$, as expressed by
\begin{equation}
 b_R(E_0;0)=\partial_E b_R(E_0;0)=0,\qquad
 \partial_E^2b_R(E_0;0)\ne0.
 \label{eq:generalzero}
\end{equation}
This condition concerns the coupling amplitude, not an extremum of the waveguide energy band. The group speed remains nonzero at $E_0$. For two interfering contact contributions, a stationary relative phase is sufficient only when their relative magnitudes also have no uncanceled linear variation. Appendix~\ref{app:general} derives this requirement.

To improve transmission of a finite pulse, we split each emitter's second-order coupling zero into two simple zeros. Both emitters retain the same contact settings, so they share the same pair of zero frequencies. Let $q$ specify this static adjustment, chosen so that $\sqrt q$ is the leading displacement of either zero for small positive $q$. Thus $q$ has units of energy squared and is the squared half-separation to leading order. We consider contact settings for which
\begin{align}
 b_R(E;q)&=a(E,q)[s(E)-q],\nonumber\\
 s(E_0+x)&=x^2+O(x^3),\qquad a(E_0,0)=a_0\ne0,
 \label{eq:realunfold}
\end{align}
Here $s(E)$ is real at real energies, $q$ is real, and the functions are analytic in a fixed neighborhood of $E_0$ and $q=0$. The zeros satisfy $s(E)=q$, so small positive $q$ gives $x_\pm=\pm\sqrt q+O(q)$. Higher-order terms can shift their midpoint. Equation~(\ref{eq:realunfold}) ensures that the two zeros move along the real energy axis. A general perturbation of a complex coupling need not have this property and can instead move them to complex energies. The physical phases and strengths realizing $q$ are held fixed while each photon propagates. Sections~\ref{sec:uniformexample} and \ref{sec:ssh} give two implementations.

The coupling to the opposite propagation direction remains finite at the carrier. Denote its rate by $\Gamma=g(E_0;0)>0$. Define the curvature coefficient $B=|a_0|^2/\Gamma$, so $u=\Gamma B(x^2-q)^2+\cdots$. The rate $g(E;q)$ and the waveguide-induced frequency shift may depend on both $E$ and $q$; $\Gamma$ denotes $g$ only at the reference point $(E_0,0)$.

Changing the contacts also shifts the emitter transition energies. We adjust the bare frequencies so that, after this waveguide-induced shift, their midpoint stays at the photon carrier. Their separation $\Delta$ is held fixed. The required bare frequencies and the resulting energy-dependent detuning are
\begin{align}
 \omega_{1,2}(q)&=E_0-\Lambda(E_0;q)\pm\Delta/2,\nonumber\\
 h(E;q)&=x-[\Lambda(E;q)-\Lambda(E_0;q)].
 \label{eq:generalcenter}
\end{align}
Thus $h(E_0;q)=0$ for every contact setting. Away from the carrier, $h$ retains the frequency dependence of the Lamb shift. The response denominators are $D_{1,2}=h\mp\Delta/2+i(\gamma+u+g)/2$. The waveguide and contact positions remain unchanged as $q$ varies.

\subsection{Cancellation between the two emitters}
The common contact amplitude determines how strongly the incident photon excites either emitter. The fields emitted by the two driven emitters must then be added in the reflected channel. For identical contacts, this sum factors into the local coupling product $c=b_L^*b_R$ and a collective interference factor $\mathcal N$. Equation~(\ref{eq:cascade}) gives
\begin{align}
 r_L&=-ic\,\mathcal N/\mathcal Q,\qquad
 \mathcal Q=D_1D_2+Pug,\nonumber\\
 \mathcal N&=(1+P)(h+i\gamma/2)
 +(1-P)[\Delta/2+i(u+g)/2].
 \label{eq:generalN}
\end{align}
Choose the separation so that the round-trip phase is an integer multiple of $2\pi$ at the carrier, $P(E_0)=1$. Its energy dependence defines $\tau$ through $P(E_0+x)=1+2i\tau x+O(x^2)$. The detuning $h$ also changes with energy; its slope is $\zeta=1-\partial_E\Lambda(E_0;0)$. At $q=\gamma=0$, expansion gives
\begin{align}
 \mathcal N&=K_0x+O(x^2),\qquad K_0=2\zeta+\Gamma\tau-i\tau\Delta,\nonumber\\
 \mathcal Q(E_0)&=-\frac{\Gamma^2+\Delta^2}{4}.
 \label{eq:generalK}
\end{align}
The term $2\zeta$ comes from the change in dressed emitter detuning, since $(1+P)h=2\zeta x+O(x^2)$. The remaining terms arise from propagation:
\begin{equation}
 (1-P)\left[\frac{\Delta}{2}+\frac{i(u+g)}{2}\right]
 =(\Gamma\tau-i\tau\Delta)x+O(x^2).
 \label{eq:collectiveslopeparts}
\end{equation}
The frequency dependence of the Lamb shift therefore contributes directly to the collective cancellation through $\zeta$.

For $K_0\ne0$, the collective factor is linear in detuning. Multiplying it by the quadratic local coupling gives a cubic reflected amplitude and a sixth-order reflection probability at zero loss. The denominator stays finite at $E_0$, so this cancellation does not require a scattering pole. Propagation matching and carrier compensation are essential. Without them, the collective factor generally remains nonzero and the quadratic local coupling gives fourth-order reflection probability. Fixed positive loss also changes the result because $\mathcal N(E_0)=i\gamma$. The special case $K_0=0$ has a higher-order collective zero and is outside the expansion considered here.

A pulse of width $\sigma$ mainly occupies detunings $x=O(\sigma)$. To keep the contact zeros in this range, their displacement must be of order $\sigma$, so $q=O(\sigma^2)$. The reflected amplitude is then of order $\sigma^3$, giving a probability of order $\sigma^6$. Each emitter excitation is of order $\sigma^2$, so independent loss contributes at order $\gamma\sigma^4$. The two probabilities compete when $\gamma=O(\sigma^2)$, with all microscopic energy scales fixed. In this joint narrowband and weak-loss limit,
\begin{align}
 R_L&=\mathcal C_R x^2(x^2-q)^2+\cdots,\nonumber\\
 A_L&=\mathcal D_\ell\gamma(x^2-q)^2+\cdots,\nonumber\\
 \mathcal C_R&=\frac{16\Gamma^2B|K_0|^2}{(\Gamma^2+\Delta^2)^2},\qquad
 \mathcal D_\ell=\frac{8\Gamma B}{\Gamma^2+\Delta^2}.
 \label{eq:generalchannels}
\end{align}
Reflection carries an extra factor $x^2$ because the two reflected fields cancel at the carrier. Independent loss has no such factor. Appendix~\ref{app:general} derives the coefficients from the frequency-dependent self-energy and controls the terms omitted in the pulse expansion.

\section{Preserving the transmitted photon pulse}
\label{sec:generalpulse}
\subsection{The desired output mode and its error}
We ask for the probability that the outgoing photon occupies the freely propagated input pulse, allowing an additional arrival delay. Reflection and absorption reduce this probability. Frequency-dependent transmission can also change the pulse shape. The mode fidelity below counts all three effects.

Let $w_\sigma(E)$ be the normalized spectral intensity of the input, centered at $E_0$, and write $\langle f\rangle=\int w_\sigma(E)f(E)\,dE$. The pulse is narrow compared with its distance from band edges and with the energy scale on which the scattering response varies. We use a fixed spectral shape rescaled by $\sigma$; before restriction to the propagating band, $\sigma$ is its standard deviation. The shape has finite moments and either compact support or rapidly decreasing tails. For the Gaussian examples,
\begin{equation}
 w_\sigma(E)=\frac{e^{-(E-E_0)^2/(2\sigma^2)}}{\sqrt{2\pi}\sigma Z_{\rm band}},
 \label{eq:weight}
\end{equation}
where $Z_{\rm band}$ is the Gaussian weight retained inside the band, and $w_\sigma$ vanishes outside it. For incidence from side $\alpha$, the overlap probability with a copy of the input pulse delayed by $t$ is
\begin{equation}
 F_\alpha(t)=\left|\int w_\sigma(E)t_\alpha(E)e^{-itx}\,dE\right|^2.
 \label{eq:F}
\end{equation}
We call this probability the mode fidelity. If $f_\sigma(E)$ is the incident spectral amplitude, the transmitted amplitude is $f_\sigma(E)t_\alpha(E)$. The target retains the same input spectral phase. Their overlap therefore contains $f_\sigma^*(E)f_\sigma(E)=w_\sigma(E)$, which explains the intensity weight in Eq.~(\ref{eq:F}). Reflection and absorption lower this unconditional fidelity because the photon is absent from the desired transmitted mode. A constant transmission phase leaves it unchanged. Adjusting $t$ corrects a linear spectral phase, while higher-order phase variation can still distort the pulse.

Flux conservation separates the mode error into loss from the transmitted channel and distortion within it. With $\widetilde t_\alpha=t_\alpha e^{-itx}$,
\begin{equation}
 1-F_\alpha(t)=\langle R_\alpha+A_\alpha\rangle+
 \left\langle\left|\widetilde t_\alpha-\langle\widetilde t_\alpha\rangle\right|^2\right\rangle.
 \label{eq:decomp}
\end{equation}
The first term counts reflected and absorbed photons. The second measures the transmitted component orthogonal to the target pulse. It can remain nonzero even when the transmitted power is close to one, because different frequencies acquire different phases. A reflection calculation must therefore be supplemented by an estimate of this mode-distortion term.

For the common contacts of Sec.~\ref{sec:general}, the exact transmission satisfies
\begin{equation}
 t_L-1=\frac{u[\gamma+(1-P)g-2ih]}{\mathcal Q}.
 \label{eq:generalt}
\end{equation}
On the occupied frequency range, $u=O(\sigma^4)$, while $h$ and $P-1$ are $O(\sigma)$. With $\gamma=O(\sigma^2)$, the deviation $t_L-1$ is therefore $O(\sigma^5)$. Its squared norm, and hence the mode-distortion term at zero delay, is $O(\sigma^{10})$. Reflection and independent loss consequently determine the leading error. Writing $\mathcal E_\alpha(t)=1-F_\alpha(t)$,
\begin{align}
 \mathcal E_L(0)={}&\mathcal C_R\langle x^2(x^2-q)^2\rangle
 +\mathcal D_\ell\gamma\langle(x^2-q)^2\rangle+o(\sigma^6).
 \label{eq:unfolding}
\end{align}
For an even spectral profile and analytic common contact controls, the next correction is $O(\sigma^8)$. The leading result refers to incidence from the weakly coupled direction. Opposite incidence can produce additional absorption and transmission-phase distortion, treated for the SSH realization in Sec.~\ref{sec:finite}.

\subsection{Zero spacing selected by the spectrum and loss}
The pulse averages depend only on the second, fourth, and sixth spectral moments. Define $M_n=\langle x^n\rangle$, with $M_0=1$. Expanding the squares in Eq.~(\ref{eq:unfolding}) gives
\begin{align}
 \mathcal E_L^{(0)}(q)={}&\mathcal C_R(M_6-2qM_4+q^2M_2)\nonumber\\
 &+\mathcal D_\ell\gamma(M_4-2qM_2+q^2).
 \label{eq:momentfunctional}
\end{align}
This expression is a quadratic function of the adjustable parameter $q$. Its derivative is
\begin{align}
 \frac{d\mathcal E_L^{(0)}}{dq}
 ={}&2(\mathcal C_RM_2+\mathcal D_\ell\gamma)q\nonumber\\
 &-2(\mathcal C_RM_4+\mathcal D_\ell\gamma M_2).
 \label{eq:momentderivative}
\end{align}
The second derivative is $2(\mathcal C_RM_2+\mathcal D_\ell\gamma)>0$ for a pulse of nonzero width. The unique leading minimum is therefore
\begin{equation}
 q_* =\frac{\mathcal C_RM_4+\mathcal D_\ell\gamma M_2}
 {\mathcal C_RM_2+\mathcal D_\ell\gamma}.
 \label{eq:generalopt}
\end{equation}
The optimum is a weighted average of the squared detuning. Define the normalized nonnegative weight
\begin{equation}
 W(E)=\frac{w_\sigma(E)(\mathcal C_Rx^2+\mathcal D_\ell\gamma)}
 {\mathcal C_RM_2+\mathcal D_\ell\gamma}.
 \label{eq:effectiveweight}
\end{equation}
We have $q_*=\int W(E)x^2\,dE$. Reflection gives extra weight to frequencies far from the carrier through its factor $x^2$ and favors $q=M_4/M_2$. Independent loss uses the original pulse weight and favors $q=M_2$. Their competition places $q_*$ between these values because $M_4\ge M_2^2$. The cost of choosing a different $q$ is explicit in
\begin{equation}
 \mathcal E_L^{(0)}(q)=\mathcal E_{L,\min}^{(0)}
 +(\mathcal C_RM_2+\mathcal D_\ell\gamma)(q-q_*)^2.
 \label{eq:completedmoment}
\end{equation}
The remaining minimum is the weighted variance of $x^2$ times $\mathcal C_RM_2+\mathcal D_\ell\gamma$. It is positive for a continuous spectrum of nonzero width: isolated zeros suppress scattering at selected frequencies, while the pulse occupies an interval. This result optimizes the common static contact family of Sec.~\ref{sec:general}. Additional controls or changes of geometry can change the leading coefficients and require a new optimization.

For a Gaussian pulse, $M_2=\sigma^2$, $M_4=3\sigma^4$, and $M_6=15\sigma^6$, up to exponentially small band-edge corrections. Define $z=q/\sigma^2$ and $\rho=\mathcal D_\ell\gamma/(\mathcal C_R\sigma^2)$. The parameter $z$ measures the squared zero displacement in units of the pulse width, while $\rho$ compares independent loss with reflection. Equation~(\ref{eq:momentfunctional}) becomes
\begin{equation}
 \frac{\mathcal E_L(0)}{\mathcal C_R\sigma^6}
 =6+(3-z)^2+\rho[2+(1-z)^2]+O(\sigma^2).
 \label{eq:balance}
\end{equation}
The reflection term is smallest at $z=3$, whereas the loss term is smallest at $z=1$. Thus loss draws the contact zeros closer to the carrier. Minimizing their sum gives
\begin{align}
 z_*&=\frac{3+\rho}{1+\rho},\nonumber\\
 \frac{\mathcal E_{L,\min}}{\mathcal C_R\sigma^6}
 &=6+2\rho+\frac{4\rho}{1+\rho}+O(\sigma^2).
 \label{eq:zopt}
\end{align}
At zero loss, the Gaussian optimum lowers the leading error coefficient from 15 to 6, giving the ratio $2/5$. This ratio depends on the pulse spectrum. For a uniform spectrum on $|x|\le\sqrt3\sigma$, the moments are $M_2=\sigma^2$, $M_4=9\sigma^4/5$, and $M_6=27\sigma^6/7$. The optimum is then $q_*=9\sigma^2/5$, and the leading error is $4/25$ of its value for coincident zeros.

For finite $\rho$, reflection and independent loss contribute at the same leading order. This limit compares static devices whose settings are chosen for progressively narrower pulses. If a nonzero $q$ is instead held fixed as the input is narrowed, the contact zeros eventually lie outside the occupied spectrum. Section~\ref{sec:fixeddevice} examines this fixed-device comparison.

Holding $\gamma>0$ fixed gives a different narrowband limit. With $q=O(\sigma^2)$, reflection and absorption both begin at fourth order and give
\begin{equation}
 \mathcal E_L(0)=\frac{8\gamma\Gamma B}{\Delta^2+(\Gamma+\gamma)^2}
 \langle(x^2-q)^2\rangle+o(\sigma^4).
 \label{eq:generalfixedloss}
\end{equation}
The optimum is $q_*=M_2$, or $z_*=1$ for a Gaussian pulse. The coefficient contains both reflected and absorbed photons and retains the full $\gamma$ dependence of the response denominator.

\subsection{A three-contact example in a uniform chain}
\label{sec:uniformexample}
The required splitting can be realized without a dimerized lattice. In an infinite uniform nearest-neighbor chain, let each emitter couple to three consecutive sites. The dispersion is $E=2J\cos k$, and a right-moving photon has spatial dependence $e^{-ikn}$ for $0<k<\pi$. Choose the relative contact couplings as $(1,-2\cos\vartheta\,e^{-ik_0},e^{-2ik_0})$, where $k_0$ is the carrier momentum and $\vartheta$ changes the middle-contact strength. For $\delta k=k-k_0$, the three excitation paths contribute the factor
\begin{align}
 &1-2\cos\vartheta\,e^{-i\delta k}+e^{-2i\delta k}\nonumber\\
 &\qquad=2e^{-i\delta k}[\cos\delta k-\cos\vartheta].
 \label{eq:uniformpaths}
\end{align}
The full coupling amplitude is this factor times $\lambda/\sqrt v$, where $\lambda$ is the common strength scale and $v=2J\sin k$. At $\vartheta=0$, the amplitude vanishes quadratically near $k_0$. A small nonzero $\vartheta$ moves the zeros to $k_0\pm\vartheta$. The carrier speed $v_0=2J\sin k_0$ is nonzero, so these momentum zeros correspond to two real frequencies. The exact control coordinate is $q=2v_0^2(1-\cos\vartheta)$, as derived in Appendix~\ref{app:uniform}.

The two three-site contact regions are displaced by $L>2$ sites. Choosing $k_0L$ to be an integer multiple of $\pi$ supplies the collective cancellation. The example $k_0=\pi/3$ and $L=6$ satisfies both the propagation and nonoverlap conditions. Coordinating the common strength and bare frequencies keeps the carrier rate difference $g-u$ and the mean shifted resonance fixed. The rates and Lamb shift away from the carrier remain frequency dependent. The resulting pulse errors approach the spectrum-dependent limits above, as illustrated in Table~\ref{tab:uniformpulse}. Appendix~\ref{app:uniform} gives the microscopic response and the pulse integration. This realization does not impose a fixed ratio $t_R/t_L$; the SSH contacts below provide that additional property.

\begin{table}[t]
\caption{Approach to the predicted narrow-pulse error in the uniform-chain realization. Entries give the left-incidence mode error at zero output delay divided by $\mathcal C_R\sigma^6$, calculated from the full frequency-dependent scattering response. For each width, the static contacts have $q=z_*\sigma^2$ and the loss rate is $\gamma=\rho\mathcal C_R\sigma^2/\mathcal D_\ell$. The two columns approach the coefficients 6 and 10 predicted by Eq.~(\ref{eq:zopt}). Parameters are $k_0=\pi/3$, $L=6$, $\Gamma=0.36J$, and $\Delta=0$.}
\label{tab:uniformpulse}
\begin{ruledtabular}
\begin{tabular}{ccc}
$\sigma/J$ & $\rho=0$, $z_*=3$ & $\rho=1$, $z_*=2$\\
$10^{-2}$ & $5.747$ & $9.233$\\
$2.5\times10^{-3}$ & $5.984$ & $9.949$\\
$6.25\times10^{-4}$ & $5.999$ & $9.997$\\
$\sigma\to0$ & $6$ & $10$
\end{tabular}
\end{ruledtabular}
\end{table}

The lossless split-to-unsplit error ratio approaches $2/5$ for the Gaussian and $4/25$ for the uniform spectrum. At $\sigma=2.5\times10^{-3}J$, the full-response ratios are $0.4000$ and $0.1601$, respectively. The two pulse shapes therefore select different zero separations and different error reductions, as predicted by their spectral moments.

\section{Realization in a dimerized waveguide}
\label{sec:ssh}
The SSH chain provides a two-contact realization of the same zero motion. A qubit couples to both sites of a unit cell, and the relative phase of the two excitation paths varies quadratically with frequency at a suitable point inside the band. Changing the contact phase separates the resulting double zero. Coordinated changes of the coupling strength and qubit frequency can also keep the ratio $t_R(E)/t_L(E)$ fixed.

\subsection{Waveguide and two-site contacts}
The waveguide is an infinite chain with two sites, A and B, in each unit cell. Photons hop between the two sites of the same cell with amplitude $t_1$ and between adjacent cells with amplitude $t_2$. Taking the cell spacing as unity, we write
\begin{equation}
 H_{\rm wg}=\sum_n\left(t_1a_n^\dagger b_n+t_2a_{n+1}^\dagger b_n+\mathrm{H.c.}\right).
 \label{eq:Hwg}
\end{equation}
Here $a_n$ and $b_n$ annihilate a photon on the A and B sites of cell $n$. Both hopping amplitudes are positive, and $J=\max(t_1,t_2)$ sets the energy unit. Qubit $j$ has transition frequency $\omega_j$ and couples to the two sites of cell $n_j$, with $n_1=0$ and $n_2=L$. Its Hamiltonian is $H_q=\sum_{j=1}^2\omega_j\sigma_j^+\sigma_j^-$, where $\sigma_j^+$ and $\sigma_j^-$ raise and lower the qubit state. We choose the phase of each qubit state so that its coupling to the B site is real and positive. The coupling to the A site then carries the relative phase $\phi_j$:
\begin{equation}
 H_{\rm int}=\sum_{j=1}^2\left[\left(\lambda_{A,j}e^{i\phi_j}a_{n_j}^\dagger+\lambda_{B,j}b_{n_j}^\dagger\right)\sigma_j^-+\mathrm{H.c.}\right].
 \label{eq:Hint}
\end{equation}
The strengths $\lambda_{A,j}$ and $\lambda_{B,j}$ are positive. Only their relative phase affects interference between the two excitation paths; changing the overall phase convention of a qubit state leaves the contact positions and physical interaction unchanged.

We use the positive band, $|t_1-t_2|<E<t_1+t_2$. A propagating photon has amplitudes on both sublattices. Their relative phase $\theta(k)$ controls interference at a qubit, while the group speed $v(k)$ sets the conversion from mode amplitude to incident flux. For $0<k<\pi$,
\begin{align}
 E(k)&=\sqrt{t_1^2+t_2^2+2t_1t_2\cos k},\nonumber\\
 \theta(k)&=\arg\left(t_1+t_2e^{ik}\right),\qquad
 v=\frac{t_1t_2\sin k}{E}>0.
 \label{eq:band}
\end{align}
Our right-moving wave has cell dependence $e^{-ikn}$, so $dk/dE=-1/v$. For left incidence, this incoming wave drives qubit $j$ through $b_{jR}$; the transmission is denoted $t_L$ because its subscript labels the side of incidence. With unit-flux normalization, the two contact paths give
\begin{equation}
 b_{jR,L}=\frac{\lambda_{A,j}e^{-i\phi_j}+\lambda_{B,j}e^{\mp i\theta}}{\sqrt{2v}},
 \quad u_j=|b_{jR}|^2,\quad g_j=|b_{jL}|^2.
 \label{eq:b}
\end{equation}
The numerator contains the amplitudes for exciting the qubit through its A and B contacts. Their relative phase determines whether they add or cancel. The factor $1/\sqrt2$ comes from the equal sublattice weights of the Bloch mode, and $1/\sqrt v$ converts its amplitude to unit flux. Thus the group speed affects the coupling rates, while cancellation is controlled by the numerator.

The waveguide shifts the qubit resonance and allows it to radiate. Including these effects, the response denominator is
\begin{equation}
 D_j=E-\omega_j+\frac{\lambda_{A,j}\lambda_{B,j}}{t_1}\cos\phi_j
       +\frac{i}{2}\left(\gamma+u_j+g_j\right).
 \label{eq:D}
\end{equation}
Its real part is the detuning from the shifted qubit resonance. Its imaginary part is half the total decay rate, including waveguide emission and independent loss. For fixed contacts, the frequency shift $-\lambda_{A,j}\lambda_{B,j}\cos\phi_j/t_1$ is constant across the band. Equation~(\ref{eq:single}) gives the local scattering amplitudes, and $p=e^{-ikL}$ accounts for propagation between the qubits. Appendix~\ref{app:green} derives these expressions from the lattice Green function.

\subsection{Why contact cancellation persists near the carrier}
\label{sec:sshstationary}
Equal contact strengths allow the two excitation paths to cancel at the carrier. Set $\lambda_{A,j}=\lambda_{B,j}=\lambda$ and $\phi_j=\phi_0=\pi+\beta$. The extra phase $\pi$ makes the paths interfere destructively when $\theta(E)=\beta$:
\begin{equation}
 u=\frac{2\lambda^2}{v}\sin^2\left(\frac{\theta-\beta}{2}\right),\qquad
 g=\frac{2\lambda^2}{v}\sin^2\left(\frac{\theta+\beta}{2}\right).
 \label{eq:rates}
\end{equation}
The ratio $\eta=u/g$ compares coupling to the incoming right-moving channel with coupling to the opposite channel. At a cancellation frequency $E_0$, $\theta(E_0)=\beta$ and $u(E_0)=0$. If $\theta$ changes linearly with $x=E-E_0$, the uncanceled amplitude is also linear in $x$. Consequently $\eta=C_2x^2+O(x^3)$, where $C_2=\theta_E(E_0)^2/(4\sin^2\beta)$ and $\theta_E=d\theta/dE$. Reflection also depends on the response denominator in Eq.~(\ref{eq:D}).

For $t_1>t_2$, the phase derivative vanishes at an interior frequency. Equation~(\ref{eq:band}) gives
\begin{equation}
 \theta_E=-\frac{E^2+t_2^2-t_1^2}{2E^2v}.
 \label{eq:thetaE}
\end{equation}
The complex number $t_1+t_2e^{ik}$ traces a circle of radius $t_2$ centered at $t_1$ on the real axis. For $t_1>t_2$, the origin lies outside the circle. At the point where a ray from the origin is tangent to the circle, the phase $\theta$ has zero first derivative, so its leading change is quadratic in detuning. The tangent triangle determines the stationary phase and energy. Swapping the bond strengths puts the origin inside the circle and removes this tangent point.

The stationary point satisfies
\begin{align}
 E_*&=\sqrt{t_1^2-t_2^2},&\cos k_*&=-t_2/t_1,\nonumber\\
 \theta_*&=\arcsin(t_2/t_1),&v_*&=t_2,
 \label{eq:stationary}
\end{align}
with $\theta_{EE}(E_*)=-1/(t_2E_*)$. Choosing the fixed phase $\beta=\theta_*$ then gives
\begin{equation}
 \eta=C_4x^4+O(x^5),\qquad
 C_4=\frac{t_1^2}{16t_2^4(t_1^2-t_2^2)}.
 \label{eq:eta4}
\end{equation}
The phase mismatch starts at $x^2$, so each qubit's incoming coupling amplitude has a second-order zero at $E_0$, while the rate ratio starts at $x^4$. We call this common contact setting the unsplit device. The coordinated changes in Sec.~\ref{sec:shaping} separate each double zero into two simple zeros. The shifted qubit response remains finite at $E_0$.

The contact coupling shifts each qubit resonance. We compensate this shift by choosing the mean bare frequency
\begin{equation}
 \omega_0=E_0+\lambda^2\cos\phi_0/t_1.
 \label{eq:omega0}
\end{equation}
The mean shifted resonance then lies at $E_0=E_*$, and its detuning is $x$. The qubit still couples to the opposite-moving channel with a finite rate,
\begin{equation}
 \Gamma_0=g(E_0)=2\lambda^2t_2/t_1^2.
 \label{eq:Gamma}
\end{equation}
For one resonant qubit at zero loss, $D(0)=i\Gamma_0/2$. Since $R_1=ug/|D|^2$, the reflection probability is $R_1=4C_4x^4+O(x^5)$. The fourth-order probability follows from the second-order coupling amplitude because the response denominator is nonzero.

At the coupling zero, an incident photon can pass through without being reflected. This is a cancellation in a scattering amplitude. A bound state instead describes an excitation localized around the emitter \cite{Calajo2016,ZhangCPL2026}.

\begin{figure*}[t]
\centering\includegraphics[width=\textwidth]{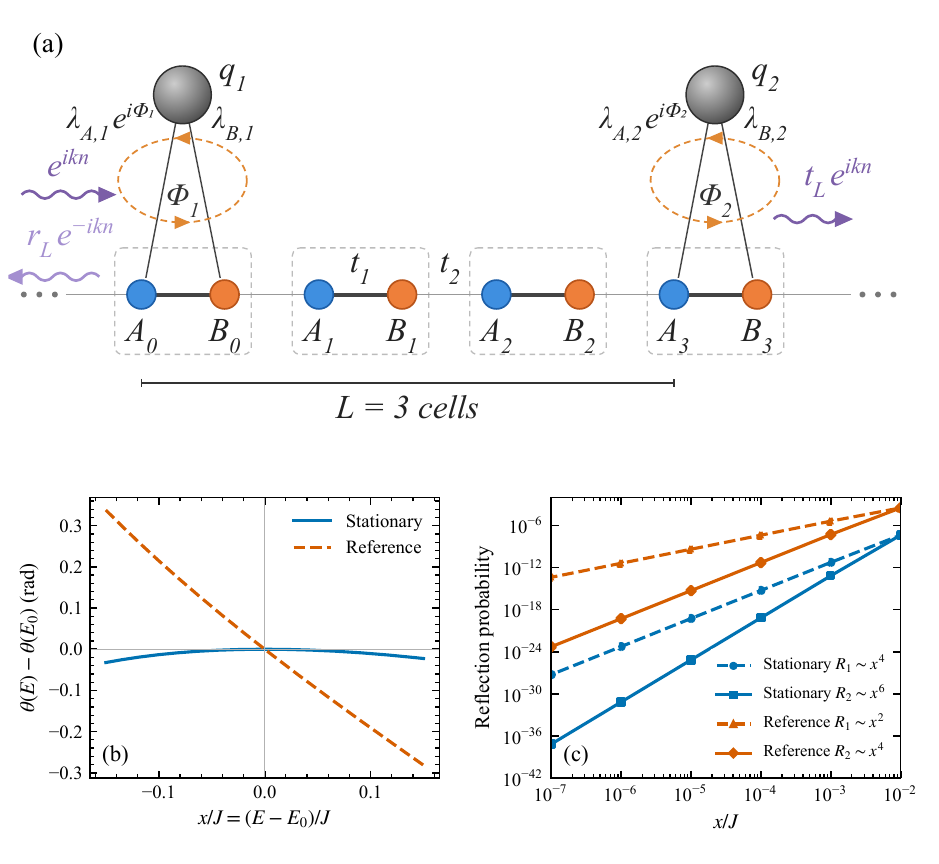}
\caption{Two stages of reflection cancellation in the SSH waveguide. (a) Two qubits couple to the A and B sites of an infinite chain and are separated by $L=3$ cells. The reference devices exchange $t_1$ and $t_2$ without moving the contacts. (b) Frequency dependence of the relative sublattice phase near $E_0=\sqrt3J/2$. For $(t_1,t_2)=(J,J/2)$, its first derivative vanishes at $E_0$; the bond-swapped chain has a nonzero slope. (c) One-qubit and two-qubit reflection probabilities $R_1$ and $R_2$ at $\Delta=\gamma=0$, calculated from Eqs.~(\ref{eq:single}) and (\ref{eq:cascade}). In the unsplit device, denoted Stationary in the plot, they scale as $x^4$ and $x^6$, whereas the reference scales as $x^2$ and $x^4$. Both devices have carrier rate $\Gamma_0=4\times10^{-2}J$, obtained with $\lambda=0.2J$ and $0.1J$, respectively. The smaller coupling used here resolves the frequency powers; subsequent pulse comparisons use $\Gamma_0=0.36J$.}
\label{fig:mechanism}
\end{figure*}

The two qubits provide a second cancellation. At zero loss, when $k_0L=n\pi$ and the mean shifted resonance is at $E_0$, their reflected fields cancel at the carrier and acquire one additional factor of $x$ nearby. For $t_2/t_1=1/2$ and $L=3$, this condition coincides with the local phase stationary point. The reflected amplitude is then cubic and its probability is sixth order. At fixed positive loss, the reflected fields no longer cancel exactly: the factor $\mathcal N$ equals $i\gamma$ at the carrier, so the probability returns to fourth order sufficiently close to $E_0$. Appendix~\ref{app:order} gives the coefficient for arbitrary symmetric qubit-frequency splitting.

The pulse calculations use $t_1=J$, $t_2=J/2$, $L=3$, and $\lambda=0.6J$ for the unsplit device. Its carrier is $E_0=\sqrt3J/2$, with $k_0=2\pi/3$, $\phi_0=7\pi/6$, $\Gamma_0=0.36J$, and $\tau=L/v_0=6/J$. To isolate the role of the sublattice phase, we compare it with devices in which $t_1$ and $t_2$ are exchanged while the contact positions stay fixed. The photon dispersion is unchanged by this exchange. One bond-swapped reference uses $\lambda=0.3J$ to keep the carrier coupling rate $\Gamma_0$ the same. The other uses $\lambda=0.6J$ to keep the contact strength the same, giving $\Gamma_{0,c}=1.44J$. Both references have $\phi_0=3\pi/2$ and their own frequency compensation from Eq.~(\ref{eq:omega0}).

\subsection{Moving the zeros at fixed relative transmission}
\label{sec:shaping}
Changing the contacts can improve the left-incidence pulse while also changing the directional response. We therefore ask whether the ratio $t_R(E)/t_L(E)$ can remain fixed. This is a ratio of complex transmission amplitudes: it contains both their amplitude ratio and their phase difference. For identical contacts, the common repeated-scattering denominator cancels, giving
\begin{equation}
 \frac{t_R}{t_L}=\prod_{s=\pm1}
 \frac{h+s\Delta/2+i[\gamma-G(E;q)]/2}
 {h+s\Delta/2+i[\gamma+G(E;q)]/2},\quad G=g-u.
 \label{eq:generalratiofunction}
\end{equation}
Here $G=g-u$ is the difference between the two local radiative rates. Keeping this ratio fixed still allows both transmission amplitudes to acquire the same frequency-dependent factor, so their individual mode errors can change. The pulse optimum in Eq.~(\ref{eq:generalopt}) does not require this additional constraint. In general, compensating the frequency shift at $E_0$ alone does not fix $h(E;q)$ at every energy, as the uniform-chain example illustrates.

For the SSH contacts, the identity $\sin\theta/v=1/t_1$ gives
\begin{equation}
 G_j\equiv g_j-u_j=-\frac{2\lambda_{A,j}\lambda_{B,j}\sin\phi_j}{t_1}.
 \label{eq:G}
\end{equation}
The two interference terms differ by $-2\sin\phi_j\sin\theta$, and division by $v$ removes the frequency dependence. This cancellation uses both the photon dispersion and the relative phase of its A and B amplitudes. Thus $G_j$ is constant for fixed contacts. The total waveguide decay rate, $g_j+u_j=G_j+2u_j(E)$, still varies with frequency and enters the response denominator.

At the unsplit contacts $G_j=\Gamma_0>0$, and the ratio becomes
\begin{equation}
 \frac{t_R}{t_L}=\prod_{s=\pm1}
 \frac{x+s\Delta/2+i(\gamma-\Gamma_0)/2}
      {x+s\Delta/2+i(\gamma+\Gamma_0)/2}.
 \label{eq:ratio_nominal}
\end{equation}
With the same $\Delta$ and $\gamma$, the unsplit device and the reference with the same $\Gamma_0$ have identical $t_R/t_L$ throughout the common positive band. At zero loss its modulus is one, so it specifies only the phase difference. Their individual reflection probabilities and pulse-mode errors can nevertheless differ. The reference with the same $\lambda$ has a different $\Gamma_0$ and answers the separate question of performance at fixed contact strength.

The derivative of the transmission phase gives the delay of a sufficiently narrow pulse. At zero loss, $\partial_E\arg(t_R/t_L)$ is the difference between the two carrier delays. Contact errors change both $G_j$ and the shifted qubit resonances, so they can also change this delay difference. Appendix~\ref{sec:robustness} bounds the amplitudes and delay difference over the error region considered in Sec.~\ref{sec:calibration}.

A change of the common contact phase moves the two local zeros apart. To preserve $t_R(E)/t_L(E)$, we also adjust the contact strength and the mean qubit frequency. Set $G=\Gamma_0>0$ and choose
\begin{align}
 \beta&=\theta_*-\nu,\qquad \phi=\pi+\beta,\qquad
 \lambda^2=\frac{Gt_1}{2\sin\beta},\nonumber\\
 \omega_{1,2}&=E_0+\frac{\lambda^2\cos\phi}{t_1}\pm\frac{\Delta}{2}.
 \label{eq:family}
\end{align}
Here $\nu$ is the phase change applied to both qubits, and the two contacts of each qubit retain equal strengths. The strength adjustment keeps $g-u=G$, while the frequency adjustment compensates the waveguide-induced resonance shift. Once selected for a pulse, these parameters remain fixed during scattering.

The local coupling vanishes at the two energies
\begin{equation}
 E_\pm=t_1\cos\beta\pm\sqrt{t_2^2-t_1^2\sin^2\beta}.
 \label{eq:zeros}
\end{equation}
For small positive $\nu$, one zero lies below $E_0$ and the other above it. Their product is $E_+E_-=E_0^2$, so the exact positions are slightly asymmetric. To leading order they lie at $E_0\pm\sqrt q$. Define $a_\theta=1/(2t_2E_0)>0$ and $q=\nu/a_\theta$; then
\begin{equation}
 \theta(E_0+x)=\theta_*-a_\theta x^2+O(x^3),\qquad
 u/G=C_4(x^2-q)^2+\cdots .
 \label{eq:contactunfold}
\end{equation}
For a prescribed spectrum and loss rate, Eq.~(\ref{eq:generalopt}) selects $q$ and hence the required phase change $\nu=a_\theta q$. Equation~(\ref{eq:family}) gives the corresponding exact static settings. If $\lambda_0$ is the strength at $\nu=0$, their small-$\nu$ changes are
\begin{align}
 \frac{\lambda(q)-\lambda_0}{\lambda_0}
 &=\frac{\nu}{2}\cot\theta_*+O(\nu^2),\nonumber\\
 \omega_0(q)-\omega_0(0)
 &=-\frac{G\nu}{2\sin^2\theta_*}+O(\nu^2).
 \label{eq:sshcontrolincrements}
\end{align}
These settings move the local zeros along the real-frequency axis. At zero loss, the two-qubit cancellation remains at $E_0$ because the propagation phase and mean shifted resonance are unchanged. The denominator stays finite there. Appendix~\ref{app:optimality} relates this construction to the real-zero condition in Eq.~(\ref{eq:realunfold}).

\subsection{Optimal nearby contact settings}
For these SSH contacts, $C_6(\Delta)$ sets the leading reflected probability and $D_{\rm loss}$ sets the leading probability of emission into the independent reservoirs. They are the coefficients $\mathcal C_R$ and $\mathcal D_\ell$ of Sec.~\ref{sec:generalpulse}, with $\Gamma=G=\Gamma_0$, $B=C_4$, and $\zeta=1$:
\begin{align}
 C_6(\Delta)&=\frac{16G^2C_4[(2+G\tau)^2+\tau^2\Delta^2]}{(G^2+\Delta^2)^2},\nonumber\\
 D_{\rm loss}&=\frac{8GC_4}{G^2+\Delta^2}.
 \label{eq:sshcoefficients}
\end{align}
The spectrum-dependent optimum therefore applies while $t_R(E)/t_L(E)$ remains fixed. We now ask whether different adjustments of the two qubits, or unequal A/B contact strengths, can do better. Fix the geometry, the carrier, and the loss rate, and allow nearby contacts with positive $G_j$. Equality of the full transmission ratio fixes $G_1=G_2=G$ and the shifted qubit frequencies to $E_0\pm\Delta/2$, up to interchange. Appendix~\ref{app:contactcoordinates} obtains these restrictions by matching the poles of the ratio.

A phase change and a strength imbalance change the local coupling in different complex directions. Let $\phi_j=\phi_0-\nu_j$, and define $r_j=\tfrac12\log(\lambda_{A,j}/\lambda_{B,j})$. The leading coupling product is proportional to $x^2-Q_j$, where
\begin{equation}
 Q_j=\frac{\nu_j-2ir_j}{a_\theta},\qquad
 \overline Q=\frac{Q_1+Q_2}{2}.
 \label{eq:maincontactcoordinates}
\end{equation}
A phase adjustment changes the real part of $Q_j$; unequal strengths add an imaginary part. To retain an error of order $\sigma^6$ when $\gamma=O(\sigma^2)$, nearby contacts must satisfy $\overline Q=O(\sigma^2)$ and $Q_1-Q_2=O(\sigma^3)$. The second restriction is stronger because different contacts produce a reflected field that is constant at leading order. Appendix~\ref{app:contactscales} proves these necessary scales from the exact response.

Write $y=x/\sigma$, $K=2+G\tau-i\tau\Delta$, and $A=(\Delta+iG)/2$. The two dimensionless contact coordinates are
\begin{equation}
 Z_{\rm c}=\frac{\overline Q}{\sigma^2},\qquad
 d_{\rm c}=\frac{A}{K}\frac{Q_1-Q_2}{\sigma^3}.
 \label{eq:scaledcontactcoordinates}
\end{equation}
The first measures the common adjustment, while the second measures the remaining difference between the qubits. After removal of a constant phase, the reflected field divided by $\sqrt{C_6}\sigma^3$ tends to
\begin{equation}
 H_3(y)+(3-Z_{\rm c})H_1(y)-d_{\rm c}H_0(y),\qquad y=x/\sigma.
 \label{eq:hermite_field}
\end{equation}
Here $H_0=1$, $H_1=y$, and $H_3=y^3-3y$. Under the unit-variance Gaussian, these three functions are orthogonal and have squared norms $1,1,6$. The common adjustment changes the linear part, and the difference between contacts adds a constant. Neither changes the remaining cubic part.

Each independent loss amplitude is instead proportional to $y^2-Z_{\rm c}$. Its Gaussian squared norm is $2+|1-Z_{\rm c}|^2$. Adding the reflected probability and the two loss probabilities, with $\rho=D_{\rm loss}\gamma/(C_6\sigma^2)$, gives
\begin{align}
 \frac{\mathcal E_L(0)}{C_6\sigma^6}
 ={}&6+|3-Z_{\rm c}|^2+\rho[2+|1-Z_{\rm c}|^2]+|d_{\rm c}|^2+o(1)\nonumber\\
 ={}&6+2\rho+\frac{4\rho}{1+\rho}
 +(1+\rho)|Z_{\rm c}-z_*|^2+|d_{\rm c}|^2+o(1).
 \label{eq:allknobs}
\end{align}
The transmitted-mode distortion is higher order in this limit. The last line therefore exposes every leading cost of the extra controls. The minimum requires real $Z_{\rm c}=z_*=(3+\rho)/(1+\rho)$ and $d_{\rm c}=0$. Equal A/B strengths and the same phase adjustment at both qubits attain it. The proof concerns nearby contacts at fixed $t_R/t_L$; it leaves finite-width corrections and subleading adjustments open.

For fixed positive loss, reflection and absorption instead give a fourth-order leading error. With common balanced contacts and $q=z\sigma^2$,
\begin{equation}
 \mathcal E_L(0)=B_\gamma\sigma^4[2+(1-z)^2]+o(\sigma^4),\quad
 B_\gamma=\frac{8\gamma GC_4}{\Delta^2+(G+\gamma)^2}.
 \label{eq:fixedloss_shared}
\end{equation}
The minimum is at $z=1$. Appendix~\ref{app:fixedloss} includes unequal contacts by adding the three outgoing probabilities as a positive quadratic form. It shows that these extra controls cannot lower the leading minimum in the fixed-loss limit either.

\section{Finite-bandwidth transmission}
\label{sec:finite}
Finite-bandwidth pulses also sample transmission-phase dispersion and frequency dependence beyond the leading expansion. These effects determine how much of the zero-splitting benefit survives for a fixed SSH device.

\subsection{Transmission phase and incidence direction}
\label{sec:sshpulse}
For left incidence at zero loss, reflection is the leading error of the unsplit device. Equation~(\ref{eq:generalt}) gives $t_L-1=O(x^5)$ because $h=x$, $u=O(x^4)$, $P-1=O(x)$, and the carrier denominator is nonzero. The transmitted-mode distortion therefore contributes at order $\sigma^{10}$, whereas reflection contributes $15C_6(\Delta)\sigma^6$. The coefficient 15 is the sixth moment of a normalized Gaussian; fixed band edges give exponentially small corrections. Right incidence has an additional frequency-dependent transmission phase. Equation~(\ref{eq:ratio_nominal}) gives
\begin{align}
 \arg(t_R/t_L)&=t_dx+b_3x^3+O(x^5),\nonumber\\
 t_d&=\frac{8\Gamma_0}{\Gamma_0^2+\Delta^2},\qquad
 b_3=-\frac{32\Gamma_0(\Gamma_0^2-3\Delta^2)}{3(\Gamma_0^2+\Delta^2)^3}.
 \label{eq:phase}
\end{align}
A time delay removes the linear phase $t_dx$ but leaves the cubic term $b_3x^3$. Its leading contribution to the Gaussian mode error is $15b_3^2\sigma^6$. For a pulse of finite width, a slightly different delay removes part of this cubic phase as well. Minimizing the variance of $b_3x^3-(t-t_d)x$ gives $t_{\rm adj}=t_d+3b_3\sigma^2+O(\sigma^4)$, and hence
\begin{align}
 1-F_L(0)&=15C_6(\Delta)\sigma^6+O(\sigma^8),\nonumber\\
 1-F_R(t_{\rm adj})&=\left[15C_6(\Delta)+6b_3^2\right]\sigma^6+O(\sigma^8).
 \label{eq:pulse6}
\end{align}
Appendix~\ref{app:pulse} shows the Gaussian averages. This local optimum approaches $t_d$ as $\sigma\to0$. The delay adjustment usually lowers the coefficient of the sixth-order error. At $|\Delta|=\Gamma_0/\sqrt3$, $b_3$ vanishes, but the positive reflection term remains. Changing $\Delta$ to reach this point also changes $t_R/t_L$, so it is a different comparison from varying the contacts at fixed Eq.~(\ref{eq:ratio_nominal}).

\begin{figure*}[t]
\centering\includegraphics[width=\textwidth]{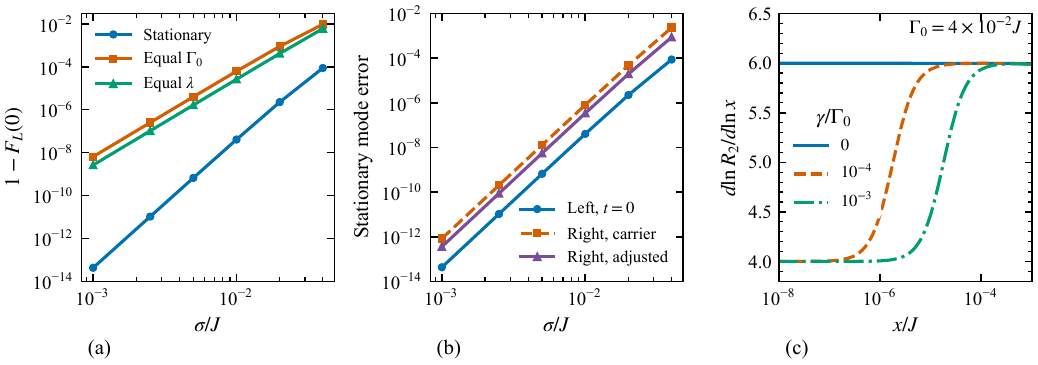}
\caption{Transmission of finite pulses and the effect of independent loss. (a) Left-incidence mode error $1-F_L(0)$ of the unsplit device and the two bond-swapped references at $\Delta=\gamma=0$. The label Stationary denotes the unsplit device of Sec.~\ref{sec:sshstationary}. (b) Errors of the same unsplit device for left incidence at zero delay and for right incidence with either the carrier delay $t_d$ or the pulse-dependent delay $t_{\rm adj}$ defined in Sec.~\ref{sec:sshpulse}. Markers in (a,b) are integrals of the full scattering response, with lines connecting the evaluated widths. (c) Local frequency exponent $d\ln R_2/d\ln x$ of the two-qubit reflection probability at $\Delta=0$. A weaker coupling, $\lambda=0.2J$ and $\Gamma_0=4\times10^{-2}J$, resolves the crossover. At every fixed positive $\gamma$, the exponent approaches four sufficiently close to the carrier, even when a sixth-order interval remains farther away. This panel concerns reflection alone. Unless stated otherwise, $t_1=J$, $t_2=J/2$, $L=3$, and $\Gamma_0=0.36J$.}
\label{fig:pulse}
\end{figure*}

At $\sigma/J=10^{-2}$ and $\Delta=\gamma=0$, the unsplit device has a left-mode error of $4.054\times10^{-8}$, almost entirely due to reflection. Both bond-swapped references reflect more than $2.6\times10^{-5}$ of the incident pulse [Fig.~\ref{fig:pulse}(a)]. The same device has a larger error for right incidence because its transmission phase varies across the pulse. Adjusting the delay reduces that error from $8.044\times10^{-7}$ to $3.427\times10^{-7}$ [Fig.~\ref{fig:pulse}(b)].

\subsection{Splitting the zeros for a prescribed bandwidth}
\begin{figure*}[t]
\centering\includegraphics[width=\textwidth]{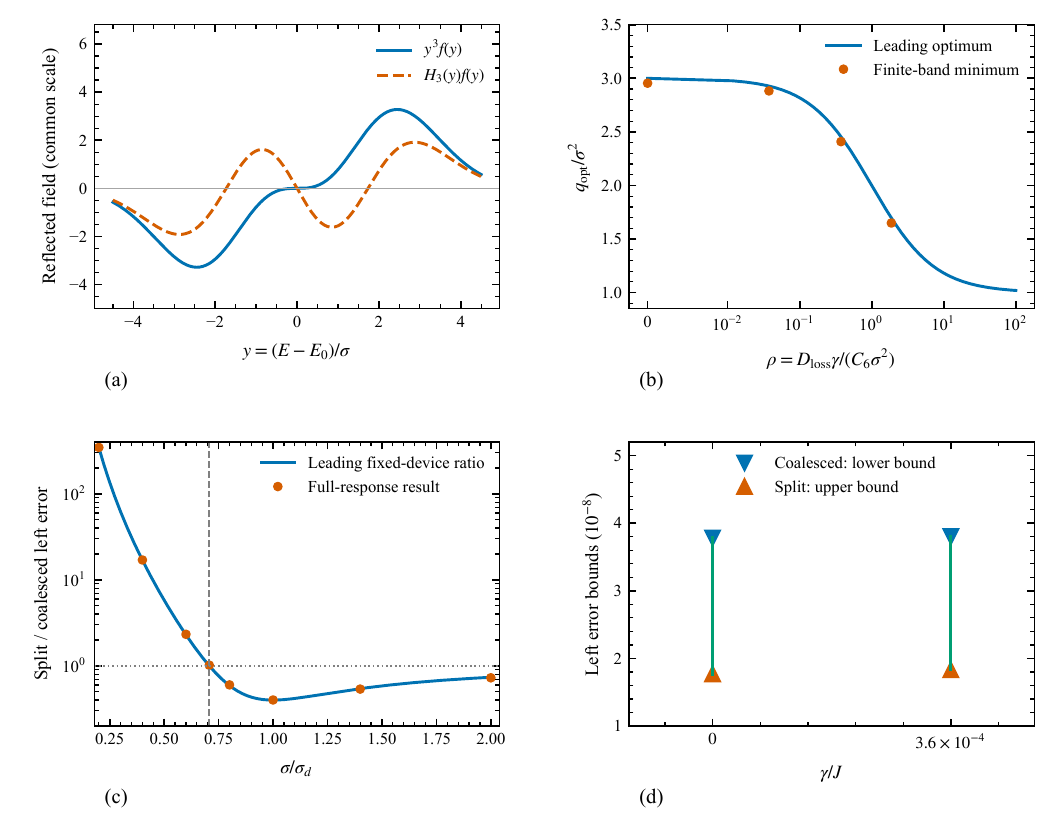}
\caption{Moving the contact zeros reduces finite-pulse error while keeping $t_R/t_L$ fixed. (a) Leading reflected spectral fields for an unsplit device and the lossless Gaussian optimum, on the same amplitude scale. With $y=(E-E_0)/\sigma$ and $f(y)=e^{-y^2/4}$, the two fields are $y^3f(y)$ and $(y^3-3y)f(y)$, respectively; $H_3(y)=y^3-3y$. (b) Optimal $q/\sigma^2$ as independent loss increases. The line is the leading prediction, and markers are local minima of the full-response pulse integral at $\sigma=10^{-2}J$. (c) Zero-delay left-mode error of one fixed split device divided by that of the unsplit device at $\gamma=0$. The split device is designed for $\sigma_d=10^{-2}J$ and held fixed as the incident width changes. The dashed vertical line marks the leading crossing estimate $\sigma/\sigma_d=1/\sqrt2$. (d) Upper bounds on the split error at zero delay and lower bounds on the unsplit error at every delay, for $\sigma=\sigma_d=10^{-2}J$ at the two indicated loss rates. Coalesced in the plot denotes unsplit zeros. All panels use $t_1=J$, $t_2=J/2$, $L=3$, $G=0.36J$, and $\Delta=0$; panels (c,d) use $q=3\sigma_d^2$.}
\label{fig:shaping}
\end{figure*}

Splitting the contact zeros changes the leading reflected field from $x^3$ to $x(x^2-q)$. For a lossless Gaussian pulse, $q=3\sigma^2$ reduces its integrated squared amplitude [Fig.~\ref{fig:shaping}(a)]. Independent loss favors a smaller separation, so the optimum moves toward the carrier as $\gamma$ increases [Fig.~\ref{fig:shaping}(b)]. Numerical integration of the exact scattering amplitudes over the pulse spectrum follows this trend. Each comparison uses contacts with the same function $t_R(E)/t_L(E)$.

Choose a design pulse with $\sigma_d=10^{-2}J$, $G=0.36J$, and $\Delta=0$. The leading lossless optimum is $q=3\sigma_d^2$, which requires a phase change $\nu\simeq3.464\times10^{-4}$ rad. To keep $t_R/t_L$ fixed, Eq.~(\ref{eq:family}) also increases the contact strength by about $1.80\times10^{-4}J$ and lowers the mean bare qubit frequency by $2.50\times10^{-4}J$.

At the design width, the split device has a smaller left-mode error than the unsplit device at both loss rates in Table~\ref{tab:pulse_errors}. The table uses the same two contact settings for $\gamma=0$ and $\gamma=3.6\times10^{-4}J$. At zero loss, the local minimum from the full response occurs at $z\simeq2.955$, close to the leading prediction $z=3$. Separating the local coupling zeros therefore improves the finite-pulse error of a device that already has sixth-order reflection at zero loss.

\begin{table}[!htbp]
\caption{Left-incidence mode error $1-F_L(0)$ at the design width $\sigma=\sigma_d=10^{-2}J$ and zero output delay. The unsplit device has $\nu=0$, and the split device has $q=3\sigma_d^2$. These two contact settings are unchanged when the loss rate is varied. The remaining parameters are $t_1=J$, $t_2=J/2$, $L=3$, $G=0.36J$, and $\Delta=0$. Values are obtained by integrating the full scattering response.}
\label{tab:pulse_errors}
\begin{ruledtabular}
\begin{tabular}{ccc}
$\gamma/J$ & Unsplit zeros & Split zeros\\
$0$ & $4.054\times10^{-8}$ & $1.628\times10^{-8}$\\
$3.6\times10^{-4}$ & $4.074\times10^{-8}$ & $1.687\times10^{-8}$
\end{tabular}
\end{ruledtabular}
\end{table}

Changing the output delay can correct spectral phase, but it cannot recover reflected or absorbed photons. The unsplit error is therefore at least $\langle R+A\rangle$ at every delay. For the split device at zero delay, Eq.~(\ref{eq:decomp}) gives the upper bound $\langle R+A+|t_L-1|^2\rangle$. The split upper bound lies below the unsplit lower bound at both loss rates, even after the finite band and Gaussian tails are included [Fig.~\ref{fig:shaping}(d)]. The resulting left-error reductions exceed $2.01\times10^{-8}$ at zero loss and $1.97\times10^{-8}$ at $\gamma=3.6\times10^{-4}J$.

For right incidence, absorption dominates at $\gamma=3.6\times10^{-4}J$. Even after the local delay adjustment, the mode error is about $7.944\times10^{-3}$. Splitting lowers it by $2.355\times10^{-8}$, a small change on this absorption background. For the unsplit device, the carrier absorption is
\begin{equation}
 A_R(0)=1-\left[\frac{\Delta^2+(G-\gamma)^2}{\Delta^2+(G+\gamma)^2}\right]^2.
 \label{eq:AR}
\end{equation}
The bond-swapped reference with the same contact strength has a larger radiative rate and can absorb less for right incidence. Thus an improvement of the left-mode error does not imply an improvement of power transmission from both directions.

A comparison at one common output delay gives a positive right-incidence improvement as well. Using $Jt_{\rm ref}\simeq22.15$ for both devices, the error reduction exceeds $2.16\times10^{-8}$ at zero loss and $2.10\times10^{-8}$ at $\gamma=3.6\times10^{-4}J$. Appendix~\ref{app:nominalcertificate} defines this delay and derives the bounds. Section~\ref{sec:calibration} and Appendix~\ref{sec:robustness} discuss contact errors.

\subsection{Changing the bandwidth of a fixed device}
\label{sec:fixeddevice}
The optimum above selects a contact setting for each pulse width. One can also keep a split device fixed and vary only the incident pulse. Then $q$ remains constant, so the zeros eventually lie outside the frequencies occupied by a sufficiently narrow pulse. At zero loss, the leading error ratio is
\begin{equation}
 \frac{\mathcal E_{L,\mathrm{split}}}{\mathcal E_{L,0}}
 \simeq\frac{15-6z+z^2}{15},\qquad z=q/\sigma^2.
 \label{eq:fixedhardware}
\end{equation}
A device designed with $q=3\sigma_d^2$ therefore loses its leading advantage when $\sigma<\sigma_d/\sqrt2$. For $\sigma_d=10^{-2}J$, integration of the full response gives a crossing at $\sigma\simeq7.099\times10^{-3}J$ without loss [Fig.~\ref{fig:shaping}(c)] and at $7.264\times10^{-3}J$ for $\gamma=3.6\times10^{-4}J$.

A fixed split device has $u_0=u(E_0)>0$. Without loss, its reflection is generically quadratic sufficiently near the carrier. With positive loss, define $H_\gamma=\Delta^2+(G+\gamma)^2$. The exact carrier transmission and the error remaining as the bandwidth vanishes are
\begin{equation}
 t_L(E_0)=\frac{H_\gamma}{H_\gamma+4\gamma u_0},\qquad
 \lim_{\sigma\to0}\mathcal E_L=1-\left(\frac{H_\gamma}{H_\gamma+4\gamma u_0}\right)^2.
 \label{eq:splitfloor}
\end{equation}
For the split device above, the limiting error is about $9.59\times10^{-10}$ at $\gamma=3.6\times10^{-4}J$. At zero loss, the reflected fields cancel at the carrier even while the qubits are excited. Positive loss both allows this excitation to decay into the reservoirs and removes exact collective reflection cancellation. Equation~(\ref{eq:splitfloor}) therefore contains absorption and reflection. The limit concerns the final outgoing mode after scattering is complete. For an untruncated transform-limited Gaussian, narrowing the spectrum also increases the temporal rms duration to $1/(2\sigma)$.

\section{Sensitivity to contact errors}
\label{sec:calibration}
\begin{figure*}[t]
\centering\includegraphics[width=\textwidth]{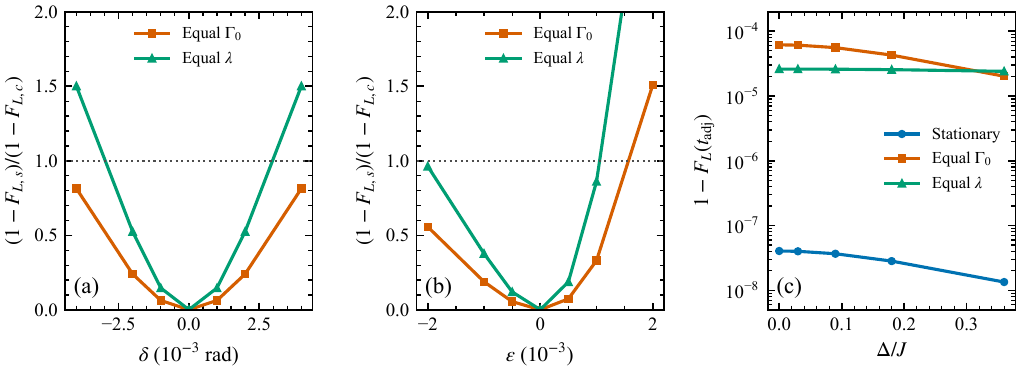}
\caption{Contact errors produce reflection at the carrier, whereas symmetric changes of the qubit frequencies preserve the lossless collective cancellation. (a,b) Ratio of the left-incidence mode error of the unsplit device (subscript $s$) to that of each bond-swapped reference (subscript $c$), at $\sigma=10^{-2}J$ and $\Delta=\gamma=0$. Each fixed device uses its locally adjusted output delay. Panel (a) varies the differential phase $\delta$ with $\epsilon=0$; panel (b) varies the A/B strength imbalance $\epsilon$ with $\delta=0$. The dotted line marks equal mode errors. Phase errors give an approximately symmetric interval of advantage. Strength imbalance can partly cancel the reference reflected field, making that interval asymmetric. (c) Left-incidence mode errors at nominal contacts when the qubit frequencies are shifted symmetrically about their fixed midpoint, at $\sigma=10^{-2}J$ and $\gamma=0$. Equation~(\ref{eq:ordercoef}) preserves sixth-order reflection for the unsplit pair at every finite fixed $\Delta$. Symbols are numerical pulse integrals and connecting lines guide the eye. The device parameters are those in Sec.~\ref{sec:sshstationary}; $s$ here denotes the unsplit stationary device, not a split device.}
\label{fig:errors}
\end{figure*}

\subsection{Which errors spoil the cancellation?}
The deliberate adjustments above include compensation of the qubit resonance shift. A contact error left uncompensated also changes the mean shifted resonance and the ratio $t_R/t_L$. If the qubits acquire different phase or strength errors, their reflected fields can fail to cancel at the carrier. With otherwise ideal contacts, a common shift of both qubit frequencies generally reduces the lossless reflection order. Opposite shifts that leave their mean fixed preserve that order.

We consider contact errors of opposite sign on the two qubits. Let $s_1=1$ and $s_2=-1$. The phase difference between the contact pairs is $\delta$, and $\epsilon$ is the difference between their A/B strength-imbalance parameters. The perturbed contacts are
\begin{align}
 \phi_j&=\phi_0+s_j\delta/2,& e_j&=s_j\epsilon/2,\nonumber\\
 \lambda_{A,j}&=\lambda(1+e_j),&\lambda_{B,j}&=\lambda(1-e_j).
 \label{eq:errors}
\end{align}
Here $\delta$ is measured in radians and $\epsilon$ is dimensionless. For $\epsilon=5\times10^{-4}$, each contact amplitude changes by $0.025\%$ from its own nominal value, with the imbalance reversed on the second qubit. This is a specified differential-error family; four arbitrary independent contact errors would require a separate comparison. The bare frequencies remain at their nominal values, so the contact-induced frequency shifts are not compensated. Appendix~\ref{app:perturbations} retains these shifts and derives the local error amplitudes.

At $\Delta=\gamma=0$, let $\theta_0$ be the nominal sublattice phase at the carrier and put $s=\sin\theta_0$. The leading reflected field is
\begin{equation}
 r_L(0)=-\frac{e^{-i\theta_0}}{s}(\delta+2i\epsilon)+O\left((|\delta|+|\epsilon|)^2\right).
 \label{eq:flooramp}
\end{equation}
Its probability is $R_L(0)=(\delta^2+4\epsilon^2)/s^2+O((|\delta|+|\epsilon|)^3)$. The coefficient is four times larger for the unsplit device, where $s=1/2$, than for a bond-swapped reference, where $s=1$. The flatter ideal frequency response is therefore more sensitive to these differential errors at the carrier.

Figure~\ref{fig:errors}(a,b) gives the resulting finite-pulse comparison. A strength imbalance can partly cancel the reference's frequency-dependent reflected amplitude because the two contributions have the same phase. A phase error contributes along the perpendicular direction in the complex-amplitude plane and does not produce that cancellation. The interval of advantage is consequently asymmetric in $\epsilon$ and approximately symmetric in $\delta$.

As the pulse narrows, its intrinsic spectral error decreases until the contact-induced carrier reflection dominates. In the plotted comparisons this sets a minimum width for an advantage over the reference. Symmetric qubit-frequency shifts behave differently [Fig.~\ref{fig:errors}(c)]. At zero loss they preserve the sixth-order power for every finite fixed $\Delta$, while changing its coefficient and the directional delay. Appendix~\ref{app:perturbations} gives the crossing values and the scales at which the local terms dominate.

\begin{figure*}[t]
\centering\includegraphics[width=\textwidth]{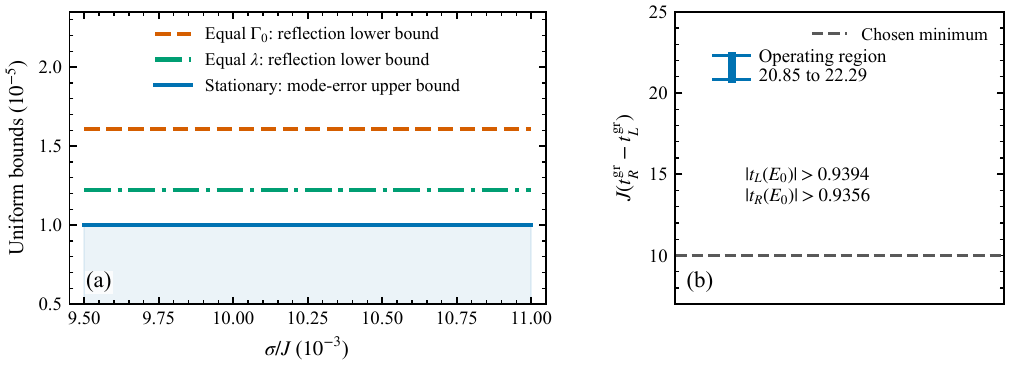}
\caption{The unsplit device retains a smaller pulse-mode error and a nonzero directional delay difference despite contact errors. (a) An upper bound on its left-mode error at zero delay lies below lower bounds on the mean reflection of both bond-swapped references. Each horizontal line applies to every contact setting and pulse width in Eq.~(\ref{eq:box}). The devices may have different errors and widths within that region, and the comparison holds for every output delay of either reference. (b) The interval $20.85/J$ to $22.29/J$ encloses the difference between the unsplit device's opposite-incidence carrier delays over the same region. It lies above the chosen minimum $10/J$. Both carrier transmission amplitudes remain nonzero, so their phase derivatives define the delays. Stationary in the plot denotes the unsplit device.}
\label{fig:joint}
\end{figure*}

\subsection{Finite-pulse performance over the specified error range}
\label{sec:continuous}
Even when contact errors spoil the exact sixth-order zero, the unsplit device can retain a smaller left-mode error than either bond-swapped reference. We allow phase error, strength imbalance, symmetric qubit-frequency splitting, and pulse width to vary simultaneously in the region
\begin{align}
 |\delta|&\le10^{-3},\qquad |\epsilon|\le5\times10^{-4},\nonumber\\
 |\Delta|/J&\le9\times10^{-2},\quad 9.5\times10^{-3}\le\sigma/J\le1.1\times10^{-2},\nonumber\\
 \gamma/J&=3.6\times10^{-4}.
 \label{eq:box}
\end{align}
The bare-frequency midpoint stays at its nominal value for each device. The comparison uses the complete scattering response, including the uncompensated shifts produced by the contact errors.

In either reference, part of the reflected field varies quadratically across the spectrum. Adding a frequency-independent field cannot cancel this variation everywhere. We isolate it by subtracting the mean of $y^2$ on a central part of the pulse. Let $y=x/\sigma$, $\varphi(y)=e^{-y^2/2}/\sqrt{2\pi}$, and retain $|y|\le T=6$, which lies inside the band throughout Eq.~(\ref{eq:box}). Define the unnormalized inner product
\begin{equation*}
 \langle f,g\rangle_T=\int_{-T}^T f^*(y)g(y)\varphi(y)\,dy,
\end{equation*}
and its moments
\begin{equation}
 \begin{gathered}
 M_n^{(T)}=\int_{-T}^Ty^n\varphi(y)\,dy,\\
 \mu_T=M_2^{(T)}/M_0^{(T)},\quad H_{2,T}(y)=y^2-\mu_T.
 \end{gathered}
 \label{eq:truncatedprojectionbasis}
\end{equation}
These truncated moments differ from the full spectral moments $M_n=\langle x^n\rangle$ of Sec.~\ref{sec:generalpulse}. By construction, $H_{2,T}$ has zero overlap with a constant or an odd field. Its squared norm is $V_T=M_4^{(T)}-[M_2^{(T)}]^2/M_0^{(T)}$. For the exact reference reflection, denoted by $r_c(\sigma y)$ with the carrier understood, Cauchy--Schwarz gives
\begin{equation}
 \int_{-T}^T|r_c(\sigma y)|^2\varphi(y)\,dy
 \ge\frac{|\langle H_{2,T},r_c\rangle_T|^2}{V_T}
 \ge\frac{[\operatorname{Re}\langle H_{2,T},r_c\rangle_T]^2}{V_T}.
 \label{eq:projection}
\end{equation}
The component along $H_{2,T}$ therefore supplies a lower bound on reflection, even if another part of the field is canceled by a contact error. Contact errors also change the frequency-dependent coefficients. Appendix~\ref{app:certificate} retains those changes and bounds the higher-order terms and the omitted spectral tails.

Let $s$ denote the unsplit device and $c$ either bond-swapped reference. Including the analytic remainders and spectral tails, the bounds are
\begin{align}
 1-F_{L,s}(0)&<1.001\times10^{-5},\nonumber\\
 \langle R_{L,c}\rangle&>1.609\times10^{-5}\quad(\text{equal }\Gamma_0),\nonumber\\
 \langle R_{L,c}\rangle&>1.222\times10^{-5}\quad(\text{equal }\lambda).
 \label{eq:certificate}
\end{align}
The unsplit device thus loses less of the target mode than either reference loses through reflection alone [Fig.~\ref{fig:joint}(a)]. These bounds hold independently over Eq.~(\ref{eq:box}), so the devices may have different contact errors and pulse widths. They also hold for every output delay of the references, since a delay cannot recover reflected photons.

The directional phase response survives in the same region. Define the carrier delay by $t_\alpha^{\rm gr}=\partial_E\arg t_\alpha|_{E_0}$, with the empty-waveguide phase removed in both directions. Both carrier transmission amplitudes remain nonzero, and
\begin{equation}
 20.85<J(t_R^{\rm gr}-t_L^{\rm gr})<22.29.
 \label{eq:responsebound}
\end{equation}
Thus the difference stays above $10/J$ [Fig.~\ref{fig:joint}(b)]. Appendix~\ref{sec:robustness} derives it from the exact transmission ratio and gives the underlying amplitude bounds.

For the chosen pulse, deliberate zero splitting improves the calibrated error by about $10^{-8}$, whereas the allowed contact errors can produce a much larger background. Appendix~\ref{app:nominalcertificate} finds a positive splitting benefit for 27 specified pairs of devices with matched errors. The continuous-region bounds above compare the unsplit device with the bond-swapped references. Thus the precision needed to resolve the extra splitting benefit is a separate question from whether the unsplit device outperforms those references.

\section{Discussion and conclusion}
\label{sec:conclusion}
For the two-emitter systems studied here, preserving a finite pulse requires choosing the cancellation frequencies for its occupied spectrum. Maximizing the cancellation order at its carrier need not minimize the total mode error. At zero loss, local contact interference gives a quadratic coupling amplitude, and cancellation between the reflected fields adds one more power of detuning. Separating the local zeros redistributes reflection across the spectrum. Independent emitter loss lacks this extra collective cancellation and therefore favors a smaller separation. The pulse spectrum sets the relative cost of moving the cancellation frequencies. For a lossless Gaussian pulse, the leading optimum is $q=3\sigma^2$, reducing the error coefficient from 15 to 6. Other pulse shapes select different spacings.

The SSH chain realizes this mechanism with an additional constraint. Its contact parameters can move the local zeros while keeping the full ratio $t_R(E)/t_L(E)$ unchanged. Within the allowed nearby contacts, equal A/B strengths and the same phase adjustment at both qubits attain the leading minimum. A uniform chain with three contacts per emitter obeys the same spectral-moment law even though its resonance shift and directional rate difference depend on frequency. The pulse optimum therefore follows from the interference mechanism rather than from the special SSH response identities.

Each optimized device has fixed settings during scattering. A device chosen for one pulse width can become unfavorable when the input is narrowed further, and fixed positive loss gives a fourth-order narrowband error for $q=O(\sigma^2)$. Transmission phase and absorption also make the error depend on the side of incidence. Contact imperfections can remove the ideal sixth-order reflection law while leaving a finite-pulse advantage. These results identify pulse spectrum, independent loss, and calibration precision as the quantities that determine a useful zero separation. Additional propagation channels, dephasing, or waveguide attenuation would require their own contributions to the mode error.

\begin{acknowledgments}
This work was supported by the National Natural Science Foundation of China (Grants No. 12675024 and No. 12275193).
\end{acknowledgments}

\section*{Data Availability}
The numerical data and code underlying the figures are available from the corresponding author upon reasonable request.

\appendix
\section{Scattering theory and pulse expansion}
\label{app:general}
\subsection{Waveguide-induced frequency shift and decay}
The waveguide changes the emitter frequency and allows it to radiate into the two propagation directions. We derive both effects from the same coupling amplitudes. For one emitter, let $a_\mu(E)$ annihilate a waveguide photon of energy $E$ propagating in direction $\mu=R,L$. We use energy-normalized operators, $[a_\mu(E),a_\nu^\dagger(E')]=\delta_{\mu\nu}\delta(E-E')$. In the one-excitation subspace, the Hamiltonian is
\begin{align}
 H={}&\omega\sigma^+\sigma^-+\sum_\mu\int dE\,E a_\mu^\dagger(E)a_\mu(E)\nonumber\\
 &+\sum_\mu\int\frac{dE}{\sqrt{2\pi}}
 [b_\mu^*(E)a_\mu^\dagger(E)\sigma^-+\mathrm{H.c.}].
 \label{eq:generalH}
\end{align}
The coupling amplitude $b_\mu(E)$ includes interference between the contact points and the waveguide density of states. For a photon incident in direction $\nu$ at energy $E$, write $e$ for the emitter excitation amplitude and $\psi_\mu(E')$ for the amplitude of a waveguide photon in direction $\mu$ at energy $E'$. Requiring outgoing waves for the scattered field gives
\begin{align}
 (E-\omega+i\gamma/2)e
 &=\sum_\mu\int\frac{dE'}{\sqrt{2\pi}}b_\mu(E')\psi_\mu(E'),\nonumber\\
 \psi_\mu(E')&=\delta_{\mu\nu}\delta(E'-E)
 +\frac{b_\mu^*(E')e}{\sqrt{2\pi}(E-E'+i0^+)}.
 \label{eq:continuumelimination}
\end{align}
Substituting the second line into the first separates the incoming drive from the self-energy. It yields $[E-\omega+i\gamma/2-\Sigma^R(E)]e=b_\nu(E)/\sqrt{2\pi}$, with
\begin{align}
 \Sigma^R(E)&=\sum_\mu\int\frac{dE'}{2\pi}
 \frac{|b_\mu(E')|^2}{E-E'+i0^+}\nonumber\\
 &=\Lambda(E)-\frac{i}{2}[u(E)+g(E)].
 \label{eq:generalself}
\end{align}
Using $(z+i0^+)^{-1}=\mathcal P(1/z)-i\pi\delta(z)$, the principal-value part of the integral gives the frequency shift $\Lambda(E)$. The delta-function part gives the decay rate $u(E)+g(E)$. Both follow from the same microscopic contact amplitudes and are not freely specified independent functions. Independent emitter loss adds $i\gamma/2$ to the inverse response. The incident state used above carries flux $1/(2\pi)$. Multiplying its amplitudes by $\sqrt{2\pi}$ gives unit incident flux and the emitter amplitude $\sqrt{2\pi}e=b_\nu/D$. The resulting scattering amplitude is $S_{\mu\nu}=\delta_{\mu\nu}-ib_\mu^*b_\nu/D$, as in Eq.~(\ref{eq:single}).

For several emitters, the waveguide also couples excitations on different emitters. The matrix $\Sigma_q^R$ contains these waveguide-mediated couplings and the individual self-energies. Let $\Omega$ be the diagonal matrix of bare emitter frequencies. The two columns of $V$ give the driving amplitudes produced by right- and left-moving incident photons. The corresponding columns of $X$ are the emitter excitation amplitudes, and $S$ maps the incident photons to the outgoing channels:
\begin{equation}
 X=(EI-\Omega-\Sigma_q^R+i\gamma I/2)^{-1}V,\qquad
 S=I-iV^\dagger X.
 \label{eq:generalGS}
\end{equation}
Completeness of the two propagating channels gives $i(\Sigma_q^R-\Sigma_q^{R\dagger})=VV^\dagger$. To see the resulting flux identity explicitly, define $K=EI-\Omega-\Sigma_q^R+i\gamma I/2$ and $\mathcal R=K^{-1}$. Then
\begin{align}
 K-K^\dagger&=i(VV^\dagger+\gamma I),\nonumber\\
 \mathcal R-\mathcal R^\dagger
 &=\mathcal R^\dagger(K^\dagger-K)\mathcal R\nonumber\\
 &=-i\mathcal R^\dagger(VV^\dagger+\gamma I)\mathcal R.
 \label{eq:resolventidentity}
\end{align}
Expansion of $S^\dagger S$ gives
\begin{align}
 S^\dagger S={}&I+iV^\dagger(\mathcal R^\dagger-\mathcal R)V\nonumber\\
 &+V^\dagger\mathcal R^\dagger VV^\dagger\mathcal R V.
\end{align}
The two terms containing $VV^\dagger$ cancel after insertion of Eq.~(\ref{eq:resolventidentity}). Since $X=\mathcal R V$, the remaining term gives
\begin{equation}
 I-S^\dagger S=\gamma X^\dagger X.
 \label{eq:generalflux}
\end{equation}
For each incident direction $\alpha$, Eq.~(\ref{eq:generalflux}) equates the missing outgoing photon probability to the loss flux divided by the incident flux. With the unit-flux normalization used here, this is $\gamma\sum_j|X_{j\alpha}|^2$. The scalar composition formula requires one further geometric condition: fields traveling between the two disjoint contact regions must be carried entirely by the two channels included above. Interleaved contacts, an additional propagating band, or an appreciable evanescent coupling between the regions require the full projected Green function instead.

\subsection{Conditions for a quadratic coupling zero}
Cancellation at one frequency fixes the relative phase of two contact contributions. To make it persist to first order in frequency, their relative magnitude must also remain unchanged. Write $b_R(E)=f_1(E)+e^{i\phi}f_2(E)$, where $\phi$ is fixed and both contributions are nonzero at $E_0$. At the carrier, $e^{i\phi}f_2(E_0)=-f_1(E_0)$. Differentiation gives
\begin{align}
 b_R'(E_0)&=f_1(E_0)\left[\frac{f_1'}{f_1}-\frac{f_2'}{f_2}\right]_{E_0}\nonumber\\
 &=f_1(E_0)\left[\partial_E\log\frac{f_1}{f_2}\right]_{E_0}.
 \label{eq:complexstationarity}
\end{align}
The first derivative vanishes when both the magnitude and the phase of $f_1/f_2$ have zero slope at $E_0$. Equal-strength SSH contacts satisfy the magnitude condition throughout the band, so only the phase slope remains. In a general mode, a changing magnitude ratio can leave a linear coupling even where the relative phase is stationary.

A smooth change of the overall Bloch-state phase multiplies $b_R$ by a nonzero phase factor. It therefore leaves both $b_R(E_0)=0$ and $b_R'(E_0)=0$ unchanged. The condition concerns the physical interference between contacts, rather than the chosen phase of the Bloch eigenvector.

The control must move the zeros to real frequencies that the pulse can occupy. A double zero alone does not guarantee this motion. For example, $a_0(x^2-iq)$ has a double zero at $q=0$, but its split zeros are complex for real $q\ne0$. Equation~(\ref{eq:realunfold}) instead makes the zero condition $s(E)=q$, with $s$ real on the real axis. Because $s$ has a nondegenerate minimum at $E_0$, a small positive $q$ gives one solution on each side. To find their positions, write $s(E_0+x)=x^2+s_3x^3+O(x^4)$ and insert $x_\pm=\pm\sqrt q+dq+O(q^{3/2})$. Both signs give $2d+s_3=0$, hence
\begin{equation}
 x_\pm=\pm\sqrt q-\frac{s_3}{2}q+O(q^{3/2}).
 \label{eq:rootmidpoint}
\end{equation}
The contact zeros have a half-separation $\sqrt q+O(q^{3/2})$ and a common shift $-s_3q/2+O(q^{3/2})$. Their midpoint can therefore move away from $E_0$. At zero loss, the cancellation between emitters remains at $E_0$ because $h(E_0;q)=0$ and $P(E_0)=1$. The mechanism does not require the two contact-zero frequencies to be exactly symmetric about the carrier.

\subsection{Reflection and loss amplitudes of two emitters}
\label{app:generalpair}
For identical contacts, reflection factors into a local coupling product and the interference between emitters. To isolate these factors, put $\Sigma=u+g$. Since $r_{1R}r_{2L}=-ug/(D_1D_2)$, the repeated-reflection denominator is $Z=\mathcal Q/(D_1D_2)$. Substitution into Eq.~(\ref{eq:cascade}) gives
\begin{equation}
 r_L=-\frac{ic}{D_1}
 -\frac{icP(D_1-iu)(D_1-ig)}{D_1\mathcal Q}.
\end{equation}
Combining the fractions first gives
\begin{equation}
 r_L=-\frac{ic[\mathcal Q+P(D_1-iu)(D_1-ig)]}{D_1\mathcal Q}.
\end{equation}
The identity
\begin{align}
 ug+(D_1-iu)(D_1-ig)
 &=ug+D_1^2-i(u+g)D_1-ug\nonumber\\
 &=D_1(D_1-i\Sigma)
\end{align}
then cancels the factor $D_1$. This reduces the reflection amplitude to $-ic[D_2+P(D_1-i\Sigma)]/\mathcal Q$. Inserting $D_{1,2}=h\mp\Delta/2+i(\gamma+\Sigma)/2$ yields Eq.~(\ref{eq:generalN}). Similarly,
\begin{align}
 &(D_1-iu)(D_2-iu)-\mathcal Q\nonumber\\
 &\quad=-iu(D_1+D_2)-u^2-Pug\nonumber\\
 &\quad=-2ihu+u(\gamma+u+g)-u^2-Pug\nonumber\\
 &\quad=u[\gamma+(1-P)g-2ih],
\end{align}
which proves Eq.~(\ref{eq:generalt}) without assuming a constant $G=g-u$ or a frequency-independent Lamb shift.

Each emitter is driven by the incident wave and by waves returned from the other emitter. The field reaching emitter 2 is $pt_{1L}/Z$. The field returned to emitter 1 is $Pr_{2L}t_{1L}/Z$. Dividing each total drive by the local response denominator gives
\begin{equation}
 q_{1L}=\frac{b_R+b_LPr_{2L}t_{1L}/Z}{D_1},\qquad
 q_{2L}=\frac{b_Rpt_{1L}}{D_2Z}.
 \label{eq:generalexcitation}
\end{equation}
For identical contacts these excitation amplitudes simplify before any expansion. Since $b_Lc=g b_R$, Eq.~(\ref{eq:generalexcitation}) becomes
\begin{align}
 q_{1L}&=\frac{b_R[D_2-iPg]}{\mathcal Q},\nonumber\\
 q_{2L}&=\frac{p b_R(D_1-iu)}{\mathcal Q}.
 \label{eq:exactexcitationreduced}
\end{align}
For example, the numerator of $q_{1L}$ contains $\mathcal Q-iPg(D_1-iu)=D_1(D_2-iPg)$; the two terms proportional to $Pug$ cancel. Therefore the exact absorption is
\begin{equation}
 A_L=\frac{\gamma u}{|\mathcal Q|^2}
 \left(|D_2-iPg|^2+|D_1-iu|^2\right),
 \label{eq:exactabsorptiongeneral}
\end{equation}
where $|p|=1$ has been used. The two terms are probabilities for emission into different loss reservoirs. They are added after taking squared moduli; no interference term joins the reservoirs.

To obtain the leading narrow-pulse loss, evaluate the smooth response factors at the unsplit, lossless carrier while retaining the small coupling $b_R$. Define $H_0=\Gamma^2+\Delta^2$ and $D_{1,2}^0=\mp\Delta/2+i\Gamma/2$. The leading emitter amplitudes are then
\begin{equation}
 q_{1L}=\frac{b_R(D_2^0-i\Gamma)}{D_1^0D_2^0}+\cdots,
 \qquad q_{2L}=\frac{p_0b_R}{D_2^0}+\cdots,
 \label{eq:generalexcitationleading}
\end{equation}
where $p_0^2=1$. Their squared norms sum to $8|b_R|^2/H_0$, because
\begin{equation}
 |D_2^0-i\Gamma|^2+|D_1^0|^2=H_0/2,\qquad
 |D_1^0D_2^0|^2=H_0^2/16.
\end{equation}
Using $|b_R|^2=\Gamma B(x^2-q)^2+\cdots$ proves $\mathcal D_\ell$ in Eq.~(\ref{eq:generalchannels}). Meanwhile $c=b_L(E_0;0)^*a_0(x^2-q)+\cdots$, with $|b_L(E_0;0)^*a_0|^2=\Gamma^2 B$. Multiplying by $\mathcal N=K_0x+\cdots$ and dividing by $\mathcal Q=-H_0/4+\cdots$ gives $\mathcal C_R$.

\subsection{Narrow-pulse error and zero separation}
We narrow the pulse without changing its spectral shape. Let $\varphi(y)\ge0$ have unit integral, zero mean, and unit variance. Before restriction to the propagating band, $w_\sigma(E_0+x)=\varphi(x/\sigma)/\sigma$. The profile may have compact support or tails decreasing faster than any inverse power. Keep the carrier a fixed distance from every channel threshold. In a fixed neighborhood of that carrier, assume analytic contact functions with uniform bounds and a denominator $\mathcal Q$ bounded away from zero for sufficiently small $q$ and $\gamma$. Introduce $x=\sigma y$, $q=z\sigma^2$, and $\gamma=\kappa_\gamma\sigma^2$. The scaled quantities $z$ and $\kappa_\gamma$ remain bounded; $\kappa_\gamma$ has units of inverse energy.

Equation~(\ref{eq:realunfold}) gives
\begin{align}
 b_R&=a_0\sigma^2(y^2-z)+O(\sigma^3),\nonumber\\
 \mathcal N&=K_0\sigma y+O(\sigma^2),\qquad
 \mathcal Q=-H_0/4+O(\sigma).
\end{align}
Writing $b_{L0}=b_L(E_0;0)$, the reflected amplitude is consequently
\begin{equation}
 r_L=\frac{4i b_{L0}^*a_0K_0}{H_0}
 \sigma^3 y(y^2-z)+O(\sigma^4).
 \label{eq:scaledreflection}
\end{equation}
For the loss channels, Eqs.~(\ref{eq:generalexcitationleading}) give
\begin{align}
 \sqrt\gamma q_{1L}
 &=\sqrt{\kappa_\gamma}a_0\frac{D_2^0-i\Gamma}{D_1^0D_2^0}
 \sigma^3(y^2-z)+O(\sigma^4),\nonumber\\
 \sqrt\gamma q_{2L}
 &=\sqrt{\kappa_\gamma}a_0\frac{p_0}{D_2^0}
 \sigma^3(y^2-z)+O(\sigma^4).
 \label{eq:scaledlossfields}
\end{align}
The extra factor $y$ in Eq.~(\ref{eq:scaledreflection}) is absent from both loss amplitudes. Squaring the reflected amplitude and the two loss amplitudes, then averaging each probability over the spectrum, gives Eq.~(\ref{eq:unfolding}). The remainders are polynomially bounded in $y$ within the expansion region, so the same orders hold after spectral averaging. Thus $r_L$ and $\sqrt\gamma q_{jL}$ are $O(\sigma^3)$, while $t_L-1$ is $O(\sigma^5)$. For a general spectral profile, the probability expansion has an $O(\sigma^7)$ remainder.

Outside the fixed carrier neighborhood, each outgoing probability is at most one. The spectral weight there is smaller than any prescribed power of $\sigma$. Restricting and renormalizing the pulse to the propagating band therefore changes none of the displayed algebraic orders.

For an even spectral profile, the $O(\sigma^7)$ probability term integrates to zero. The leading reflected amplitude is odd in $y$ and its next correction is even, so their interference term is odd. The latter correction contains $x^4$, $qx^2$, and $\gamma(x^2-q)$. In each loss channel, the leading amplitude is even and its next correction is odd, with the same consequence. This gives the $O(\sigma^8)$ remainder for analytic common contact changes. The argument uses the symmetry of the incident spectrum; it does not require an even dispersion or an even Lamb shift. Setting $\delta t=t_L-1$ in Eq.~(\ref{eq:generalt}) also gives
\begin{align}
 \operatorname{Var}(t_L)&=\langle|\delta t|^2\rangle-|\langle\delta t\rangle|^2,\nonumber\\
 0\le\operatorname{Var}(t_L)&\le\langle|\delta t|^2\rangle=O(\sigma^{10}).
\end{align}
The exact response satisfies flux conservation before expansion, so using this bound does not discard a transmission probability of the same order as reflection or absorption.

The derivative in Eq.~(\ref{eq:momentderivative}) has positive slope, which gives the unique leading optimum in Eq.~(\ref{eq:generalopt}). The minimized leading error is
\begin{align}
 \mathcal E_{L,\min}^{(0)}={}&\mathcal C_RM_6+\mathcal D_\ell\gamma M_4\nonumber\\
 &-\frac{(\mathcal C_RM_4+\mathcal D_\ell\gamma M_2)^2}
 {\mathcal C_RM_2+\mathcal D_\ell\gamma}.
 \label{eq:generalmin}
\end{align}
Since $M_4\ge M_2^2$, the optimum lies between $M_2$ and $M_4/M_2$. For a continuous spectrum of nonzero width, the remaining variance is positive. The scaled error converges uniformly on bounded intervals of $z$. Within such an interval containing the unique leading minimum, the minimizer approaches Eq.~(\ref{eq:generalopt}). Higher-order terms can shift it at finite pulse width. For an even spectral profile and analytic contact controls, the shift in $z$ is $O(\sigma^2)$ at fixed scaled loss weight. This optimization varies the common real zero separation within Eq.~(\ref{eq:realunfold}); changes of contact geometry require a different optimization.

\subsection{Fixed nonzero loss and directional transmission}
We retain $q=z\sigma^2$, with bounded $z$, while holding $\gamma>0$ fixed as the pulse narrows. The collective numerator no longer vanishes at the carrier; its value is $i\gamma$. To obtain the leading coefficients, we evaluate the smooth response factors at the reference setting $(E,q)=(E_0,0)$. Define $H_\gamma=\Delta^2+(\Gamma+\gamma)^2$ and $D_{1,2}^\gamma=\mp\Delta/2+i(\Gamma+\gamma)/2$. The carrier denominator at this reference setting is $-H_\gamma/4$, giving
\begin{align}
 R_L&=\frac{16\Gamma^2B\gamma^2}{H_\gamma^2}(x^2-q)^2+\cdots,\nonumber\\
 A_L&=\frac{8\gamma\Gamma B(\Delta^2+\Gamma^2+\gamma^2)}{H_\gamma^2}(x^2-q)^2+\cdots.
 \label{eq:generallossseparate}
\end{align}
For the absorption coefficient, Eq.~(\ref{eq:exactabsorptiongeneral}) at the carrier gives
\begin{align}
 &|D_2^\gamma-i\Gamma|^2+|D_1^\gamma|^2\nonumber\\
 &\quad=\frac{\Delta^2+(\gamma-\Gamma)^2}{4}
 +\frac{\Delta^2+(\gamma+\Gamma)^2}{4}\nonumber\\
 &\quad=\frac{\Delta^2+\Gamma^2+\gamma^2}{2}.
\end{align}
The squared denominator is $H_\gamma^2/16$. Adding the reflected and absorbed coefficients then gives
\begin{align}
 &\frac{8\gamma\Gamma B}{H_\gamma^2}
 \left[2\Gamma\gamma+\Delta^2+\Gamma^2+\gamma^2\right]\nonumber\\
 &\hspace{1cm}=\frac{8\gamma\Gamma B}{H_\gamma},
\end{align}
which proves Eq.~(\ref{eq:generalfixedloss}). Equation~(\ref{eq:generalt}) gives $t_L-1=O(\sigma^4)$ and a variance of order $\sigma^8$, which is subleading to the fourth-order error. The common displacement is minimized at $q=M_2$.

The directional transmission ratio follows from the local numerators $h\mp\Delta/2+i(\gamma+G)/2$ for left incidence and $h\mp\Delta/2+i(\gamma-G)/2$ for right incidence. Their product ratio is Eq.~(\ref{eq:generalratiofunction}). Keeping both functions $G(E;q)$ and $h(E;q)$ unchanged preserves it across the band. Compensating the carrier frequency fixes $h$ only at $E_0$. The general pulse optimum consequently allows the directional ratio to change as $q$ changes.

\subsection{Additional cancellation between the reflected fields}
\label{app:order}
\subsubsection{Carrier matching and reflection coefficient}

Without local loss, the two reflected fields can cancel at the carrier even when the qubits have different frequencies. To separate this cancellation from interference within each contact pair, keep the mean qubit frequency fixed and allow the symmetric offsets
\begin{equation}
 \omega_{1,2}=\omega_0\pm\Delta/2
 \label{eq:split}
\end{equation}
at nominal contacts. With $\Sigma=u+g$, the denominators are $D_{1,2}=x\mp\Delta/2+i(\gamma+\Sigma)/2$. The exact left reflection reduces to
\begin{align}
 r_L&=-ic\,N/Q,\qquad Q=D_1D_2+Pug,\nonumber\\
 N&=(1+P)(x+i\gamma/2)+(1-P)(\Delta/2+i\Sigma/2).
 \label{eq:N}
\end{align}
The offsets enter the two dressed detunings with opposite signs and cancel in their sum at the fixed midpoint. The general reduction in Sec.~\ref{app:generalpair}, specialized to $h=x$, gives Eq.~(\ref{eq:N}).

The propagation phase must satisfy $k_0L=n\pi$, with integer $n$. At the stationary SSH carrier this requires $L\arccos(-t_2/t_1)=n\pi$, so the bond ratio and qubit separation must be chosen together. The condition $t_1>t_2$ produces the local quadratic zero but does not by itself fix the propagation phase. The choice $t_2/t_1=1/2$ and $L=3$ satisfies both requirements. It gives $P(0)=1$ and
\begin{equation}
 N(0)=i\gamma,\qquad
 Q(0)=-\frac{\Delta^2+(\Gamma_0+\gamma)^2}{4}\ne0.
 \label{eq:carrierNQ}
\end{equation}
At zero loss, the collective numerator has an additional zero even for nonzero $\Delta$. If $\tau=L/v_0$, then $P=1+2i\tau x+O(x^2)$ and
\begin{equation}
 N=\left(2+\Gamma_0\tau-i\tau\Delta\right)x+O(x^2).
 \label{eq:Nlinear}
\end{equation}
The local quadratic amplitude and this collective linear factor give a cubic reflected amplitude. More generally, if $u/g=C_{2m}x^{2m}+\cdots$, then
\begin{align}
 R_2&=\mathcal{C}_{2m+2}(\Delta)x^{2m+2}+\cdots,\nonumber\\
 \mathcal{C}_{2m+2}(\Delta)&=\frac{16\Gamma_0^2C_{2m}\left[(2+\Gamma_0\tau)^2+\tau^2\Delta^2\right]}{(\Gamma_0^2+\Delta^2)^2}.
 \label{eq:ordercoef}
\end{align}
This coefficient is positive for every finite real $\Delta$. The unsplit pair has $m=2$ and sixth-order reflection, whereas the ordinary-contact pair has $m=1$ and fourth-order reflection. Writing $C_6(\Delta)=\mathcal{C}_6(\Delta)$ for the unsplit pair, at $\Delta=0$ we obtain $C_6=16C_4(\tau+2/\Gamma_0)^2$. Without propagation matching, the unsplit pair retains a fourth-order reflection probability at zero loss. The collective cancellation depends on the mean dressed frequency being $E_0$, while the qubits may be symmetrically detuned. A common shift of the bare frequencies generally makes $N(0)\ne0$ and restores fourth-order reflection. At fixed $\gamma>0$, $N(0)=i\gamma$ removes the collective zero, so the unsplit pair has fourth-order reflection sufficiently close to the carrier [Fig.~\ref{fig:pulse}(c)].

The central device and both bond-swapped references use the parameters of Sec.~\ref{sec:sshstationary}. The comparison keeps the attachment sites fixed while exchanging $t_1$ and $t_2$. This changes the contact amplitudes of the Bloch mode even though the dispersion is unchanged.

\subsubsection{SSH coefficients and the fixed-loss limit}
At zero loss and the carrier where the right-moving photon does not excite either qubit, $u(0)=c(0)=0$ and $g(0)=\Gamma_0$. The denominator in Eq.~(\ref{eq:carrierNQ}) is finite. The frequency dependence of reflection therefore comes from the product $cN$: if $c$ starts at $x^m$ and $N$ starts at $x$, then $r_L$ starts at $x^{m+1}$. In particular, $u/g=C_{2m}x^{2m}+\cdots$ gives $|c|^2=\Gamma_0^2C_{2m}x^{2m}+\cdots$. Using Eq.~(\ref{eq:Nlinear}) and $|Q(0)|^2=(\Gamma_0^2+\Delta^2)^2/16$ yields Eq.~(\ref{eq:ordercoef}). Since $2+\Gamma_0\tau>0$, the collective factor is linear for every finite real $\Delta$ in this geometry.

The frequency-independent rate difference gives an additional simplification. At zero loss, define $W_0=(x+i\Gamma_0/2)^2-\Delta^2/4$. Since $g=\Gamma_0+u$, the two local transmission numerators multiply to $W_0$, and the exact pair response is
\begin{equation}
 \begin{aligned}
 Q&=W_0+u[2ix+(P-1)g],\\
 t_L&=W_0/Q.
 \end{aligned}
 \label{eq:tphase}
\end{equation}
This is the SSH specialization of Eq.~(\ref{eq:generalt}). At the quadratic contact zero, $u=O(x^4)$ and $P-1=O(x)$, while $W_0(0)=-(\Gamma_0^2+\Delta^2)/4$ is finite. Hence $t_L-1=O(x^5)$, with an imaginary leading coefficient. The resulting variation of the transmitted mode enters the pulse error at higher order than reflection.

For fixed positive loss, the exact excitation fields and absorption are given by Eqs.~(\ref{eq:exactexcitationreduced}) and (\ref{eq:exactabsorptiongeneral}), with $h=x$. Their carrier values have $\Gamma=\Gamma_0$ and $p_0=e^{-ik_0L}=(-1)^n$; the latter has unit modulus and does not change absorption. Inserting $|b_R|^2=\Gamma_0 C_{2m}x^{2m}+\cdots$ gives reflection and absorption of order $x^{2m}$. For the stationary contact, $m=2$ and $C_{2m}=C_4$, the coefficients in Eq.~(\ref{eq:generallossseparate}) give $B_R$, $B_A$, and their sum $B_E$ in Eq.~(\ref{eq:losscoef}). Reflection is quadratic in weak local loss, whereas absorption is linear.

\subsection{Loss changes the asymptotic order}
Fixed nonzero local loss gives $N(0)=i\gamma$ and removes the cancellation between the reflected fields. The nominal local contact zero remains, so $r_L(0)=0$ and $t_L(0)=1$. If the local coupling amplitude starts at $x^m$, the reflected probability now starts at $x^{2m}$, rather than $x^{2m+2}$. With $H_\gamma=\Delta^2+(\Gamma_0+\gamma)^2$, expansion of the exact response gives
\begin{align}
 R_L&=B_Rx^{2m}+\cdots,&B_R&=\frac{16\Gamma_0^2\gamma^2C_{2m}}{H_\gamma^2},\nonumber\\
 1-T_L&=B_Ex^{2m}+\cdots,&B_E&=\frac{8\gamma\Gamma_0 C_{2m}}{H_\gamma},\nonumber\\
 A_L&=B_Ax^{2m}+\cdots,&B_A&=\frac{8\gamma\Gamma_0 C_{2m}(\Delta^2+\Gamma_0^2+\gamma^2)}{H_\gamma^2}.
 \label{eq:losscoef}
\end{align}
The coefficients satisfy $B_E=B_R+B_A$. Reflection is quadratic in weak loss, while absorption is linear. For the unsplit device, weak absorption can dominate the mode error before it changes the apparent reflection power over an intermediate detuning interval. A sixth-order reflection segment therefore does not by itself establish sixth-order total pulse error at fixed loss.

Writing $\mathcal K=(2+\Gamma_0\tau)^2+\tau^2\Delta^2$, the weak-loss balances with Eq.~(\ref{eq:ordercoef}) give
\begin{equation}
 x_{R,\rm cross}\simeq\frac{\gamma}{\sqrt{\mathcal K}},\qquad
 x_{E,\rm cross}\simeq\sqrt{\frac{\gamma(\Delta^2+\Gamma_0^2)}{2\Gamma_0\mathcal K}}.
 \label{eq:cross}
\end{equation}
For an unsplit Gaussian pulse, $\langle x^6\rangle/\langle x^4\rangle=5\sigma^2$. Replacing each detuning scale in Eq.~(\ref{eq:cross}) by $\sigma=x_{\rm cross}/\sqrt5$ therefore estimates where the corresponding leading pulse contributions balance. These estimates compare local asymptotic terms; the full finite-band response determines the actual crossing. Narrowing at fixed $\gamma>0$ gives a different leading power from taking $\gamma=0$ first.

\section{Microscopic waveguide realizations}
\label{app:realizations}
The two waveguides realize the same local and collective interference with different microscopic contact amplitudes.

\subsection{Three-contact realization in a uniform chain}
\label{app:uniform}
Three coupling points on a uniform chain provide a second way to split a quadratic coupling zero. The construction in Sec.~\ref{sec:uniformexample} does not require an SSH sublattice phase or a frequency-independent Lamb shift. Consider the infinite chain $H_{\rm wg}=J\sum_n(a_{n+1}^\dagger a_n+\mathrm{H.c.})$, with $J>0$. Its band is $-2J<E<2J$, with $E=2J\cos k$ for $0<k<\pi$. A right-moving wave has spatial factor $e^{-ikn}$ and speed $v=2J\sin k$. The three contact amplitudes are
\begin{equation}
 (w_0,w_1,w_2)=\lambda(1,-2\cos\vartheta\,e^{-ik_0},e^{-2ik_0}),
 \label{eq:uniformcontacts}
\end{equation}
through
\begin{equation}
 H_{\rm int}=\sum_{j=1}^{2}\sum_{n=0}^{2}
 \left(w_n a_{n_j+n}^\dagger\sigma_j^-+\mathrm{H.c.}\right).
 \label{eq:uniforminteraction}
\end{equation} Here $n_1=0$ and $n_2=L>2$, so the three-site contact regions do not overlap. The parameter $\vartheta$ changes the middle-contact strength relative to the outer two and is fixed during each pulse. Let $E_0=2J\cos k_0$, $v_0=2J\sin k_0$, and $\delta k=k-k_0$. Including the unit-flux prefactor in Eq.~(\ref{eq:uniformpaths}) gives
\begin{align}
 b_R&=\frac{\lambda}{\sqrt v}[1-2\cos\vartheta\,e^{-i\delta k}+e^{-2i\delta k}]\nonumber\\
 &=\frac{\lambda e^{-i\delta k}}{\sqrt v\,v_0^2}[q-s(E)],\nonumber\\
 q&=2v_0^2(1-\cos\vartheta),\qquad
 s(E)=2v_0^2[1-\cos\delta k].
 \label{eq:uniformfactor}
\end{align}
Near the carrier, $\delta k=-x/v_0+O(x^2)$, so $s(E)=x^2+O(x^3)$. The prefactor in Eq.~(\ref{eq:uniformfactor}) is analytic and nonzero near $E_0$, which gives Eq.~(\ref{eq:realunfold}) with $a=-\lambda e^{-i\delta k}/(\sqrt v\,v_0^2)$. For left-moving incidence, replace $\delta k$ by $-(k+k_0)$ in the first line of Eq.~(\ref{eq:uniformfactor}).

The retarded lattice Green function is
\begin{equation}
 G_{nm}^R(E)=-\frac{i}{v}e^{-ik|n-m|}.
 \label{eq:uniformgreen}
\end{equation}
The self-energy is $\Sigma^R=\sum_{n,m=0}^2w_n^*G_{nm}^Rw_m$. Group the terms by the separation of the contact sites. After extracting the common factor $-i\lambda^2/v$, the on-site terms give $2+4\cos^2\vartheta$. Pairs one site apart give $-8\cos\vartheta\cos k_0\,e^{-ik}$, and the pair two sites apart gives $2\cos(2k_0)e^{-2ik}$. Hence
\begin{align}
 \Sigma^R=-\frac{i\lambda^2}{v}\big[&2+4\cos^2\vartheta
 -8\cos\vartheta\cos k_0\,e^{-ik}\nonumber\\
 &+2\cos(2k_0)e^{-2ik}\big].
 \label{eq:uniformselfsum}
\end{align}
Taking the real part and using $v=2J\sin k$ gives the Lamb shift below. To obtain the rate difference directly, the two contact polynomials give
\begin{align}
 u&=\frac{4\lambda^2}{v}[\cos(k-k_0)-\cos\vartheta]^2,\nonumber\\
 g&=\frac{4\lambda^2}{v}[\cos(k+k_0)-\cos\vartheta]^2.
 \label{eq:uniformrates}
\end{align}
Their difference factors into $\cos(k+k_0)-\cos(k-k_0)=-2\sin k\sin k_0$ and $2\cos k\cos k_0-2\cos\vartheta$. Combining these expressions gives
\begin{align}
 \Lambda(E;q)&=\frac{4\lambda^2\cos\vartheta\cos k_0}{J}
 -\frac{\lambda^2\cos(2k_0)}{J^2}E,\nonumber\\
 G(E;q)&=\frac{8\lambda^2\sin k_0}{J}
 \left(\cos\vartheta-\frac{E\cos k_0}{2J}\right).
 \label{eq:uniformself}
\end{align}

Both $\Lambda$ and $G$ retain an energy dependence. We can nevertheless keep the carrier rate difference $G(E_0;q)$ and the mean shifted emitter frequency fixed while moving the zeros. Choose
\begin{equation}
 \lambda^2(q)=\frac{\Gamma J}{8\sin k_0(\cos\vartheta-\cos^2 k_0)}
 \label{eq:uniformstrength}
\end{equation}
for sufficiently small $q$, so that $G(E_0;q)=\Gamma$. The opposite-direction rate is $g(E_0;q)=\Gamma+u(E_0;q)$ and equals $\Gamma$ at the unsplit setting. Adjusting the bare frequencies according to Eq.~(\ref{eq:generalcenter}) also keeps the mean shifted frequency at $E_0$. The energy dependence away from the carrier remains in the response. With $\lambda_0=\lambda(0)$, the coefficients entering the general pulse law are
\begin{equation}
 B=\frac{\lambda_0^2}{\Gamma v_0^5},\qquad
 \zeta=1+\frac{\lambda_0^2\cos(2k_0)}{J^2},\qquad \tau=L/v_0.
 \label{eq:uniformcoefficients}
\end{equation}

For $k_0=\pi/3$, $L=6$, $\Gamma=0.36J$, and $\Delta=0$, the propagation is matched and the contact regions do not overlap. These parameters give $E_0=J$, $B=1/(81J^4)$, $\zeta\simeq0.9654$, $\mathcal C_R\simeq15.39/J^6$, and $\mathcal D_\ell\simeq0.2743/J^5$. Table~\ref{tab:uniformpulse} gives the mode error from the full band response as the width decreases. The limiting coefficients 6 and 10 follow from Eq.~(\ref{eq:zopt}); they do not rely on a constant Lamb shift.

The six contact sites also give a direct microscopic scattering calculation. Let $W$ be the $6\times2$ matrix whose $j$th column contains $w_0,w_1,w_2$ at the sites coupled to emitter $j$ and is zero elsewhere. Then $\Sigma_q^R=W^\dagger G^R W$ and Eq.~(\ref{eq:generalGS}) gives $S$. All contacts of emitter 1 lie to the left of those of emitter 2. The Green function between the regions therefore factors into outgoing waves, making the scalar composition exact for this geometry.

The pulse integrals use Gaussian quadrature on $|x|\le12\sigma$, with the retained Gaussian normalized to one. The omitted weight is below $3.6\times10^{-33}$. The uniform profile is integrated over its entire support. Reflection, the two independent loss probabilities, and the variance of $t_L-1$ are integrated separately to resolve the small total error.

\subsection{Microscopic scattering calculation for the SSH waveguide}
\label{app:green}
The lattice Green function gives the field produced at one contact site by excitation at another. Its projection onto the four contacts determines propagation between the qubits and their waveguide-induced frequency shifts and decay. For a real frequency in the positive band, set $z=e^{-ik}$ and $s_k=2it_1t_2\sin k$. In the site order $(A_0,B_0,A_L,B_L)$, the projected matrix is
\begin{equation}
 \mathcal{G}(E)=\begin{pmatrix}
 g_0&g_1&g_L&g_{AB}\\
 g_1&g_0&g_{BA}&g_L\\
 g_L&g_{BA}&g_0&g_1\\
 g_{AB}&g_L&g_1&g_0
 \end{pmatrix},
 \label{eq:contactG}
\end{equation}
where
\begin{align}
 g_0&=E/s_k,&g_1&=(t_1+t_2z)/s_k,\nonumber\\
 g_L&=Ez^L/s_k,&g_{AB}&=z^L(t_1+t_2z)/s_k,\nonumber\\
 g_{BA}&=z^{L-1}(t_1z+t_2)/s_k.
 \label{eq:gentries}
\end{align}
The outgoing-wave prescription and sublattice indices follow from the lattice integral
\begin{align}
 I_d(E)&=\int_{-\pi}^{\pi}\frac{dq}{2\pi}
 \frac{e^{iqd}}{(E+i0^+)^2-t_1^2-t_2^2-2t_1t_2\cos q}\nonumber\\
 &=\frac{e^{-ik|d|}}{2it_1t_2\sin k}.
 \label{eq:Id}
\end{align}
For $d\ge0$, the substitution $\zeta=e^{iq}$ gives
\begin{equation}
 I_d=-\frac{1}{t_1t_2}\oint_{|\zeta|=1}\frac{d\zeta}{2\pi i}
 \frac{\zeta^d}{(\zeta-\zeta_-)(\zeta-\zeta_+)}.
\end{equation}
The two roots satisfy $\zeta_-\zeta_+=1$ and $\zeta_-+\zeta_+=[(E+i0^+)^2-t_1^2-t_2^2]/(t_1t_2)$. In the positive band, the retarded prescription places $\zeta_-\to e^{-ik}$ inside the unit circle and $\zeta_+\to e^{ik}$ outside. The enclosed residue is $-e^{-ikd}/[t_1t_2(e^{-ik}-e^{ik})]$, which gives Eq.~(\ref{eq:Id}). Since the denominator is even in momentum, $I_{-d}=I_d$.

With row site $\alpha_n$, column site $\beta_m$, and $d=n-m$, the matrix elements are
\begin{align}
 G_{A_n,A_m}&=G_{B_n,B_m}=E I_d,\nonumber\\
 G_{A_n,B_m}&=t_1I_d+t_2I_{d-1},\nonumber\\
 G_{B_n,A_m}&=t_1I_d+t_2I_{d+1}.
 \label{eq:Gindex}
\end{align}
For the off-diagonal contact entries, $g_{AB}=G_{A_0,B_L}$ contains $I_{-L-1}$, while $g_{BA}=G_{B_0,A_L}$ contains $I_{-L+1}$. Substitution gives Eq.~(\ref{eq:gentries}). The root $z=e^{-ik}$ describes outgoing propagation in our convention. Both propagating channels contribute to the imaginary part; no finite artificial broadening is introduced inside the band.

The matrix $W$ records how each qubit couples to the four contact sites:
\begin{equation}
 W=\begin{pmatrix}
 \lambda_{A,1}e^{i\phi_1}&0\\\lambda_{B,1}&0\\
 0&\lambda_{A,2}e^{i\phi_2}\\0&\lambda_{B,2}
 \end{pmatrix}.
\end{equation}
To describe an incident photon, evaluate its normalized Bloch wave at the same sites. The right- and left-moving site amplitudes are $\chi_R$ and $\chi_L$. Contracting them with $W$ and dividing by $\sqrt v$ gives the unit-flux driving columns of $V$:
\begin{equation}
 \chi_R=\frac{1}{\sqrt2}\begin{pmatrix}1\\e^{-i\theta}\\e^{-ikL}\\e^{-i(\theta+kL)}\end{pmatrix},
 \qquad\chi_L=\chi_R^*,\qquad
 V=\frac{W^\dagger(\chi_R,\chi_L)}{\sqrt v}.
\end{equation}
The waveguide contribution to the qubit response is $\Sigma_q=W^\dagger\mathcal G W$. Solving the resulting two-qubit response gives the excitation amplitudes $X$:
\begin{equation}
 X=\left[E I-\operatorname{diag}(\omega_1,\omega_2)+i\gamma I/2-\Sigma_q\right]^{-1}V.
\end{equation}
These excitations radiate into the waveguide through $V^\dagger$ and into the independent loss channels at rate $\gamma$. The outgoing amplitudes and loss probabilities are therefore
\begin{equation}
 S=I-iV^\dagger X,\qquad A_\alpha=\gamma\sum_j|X_{j\alpha}|^2.
 \label{eq:GS}
\end{equation}
The rows of $S$ label right- and left-moving outgoing photons. Its first and second columns correspond to photons incident from the left and right. Thus $S_{11}=t_L$ and $S_{22}=t_R$; the off-diagonal entries are reflection amplitudes at the same spatial reference. The identity $i(\Sigma_q-\Sigma_q^\dagger)=VV^\dagger$ ensures that the outgoing waveguide probability plus the two loss probabilities equals one.

For one contact pair, the diagonal Green entry is $-i/(2v)$. The off-diagonal entry obeys
\begin{equation}
 \frac{t_1+t_2e^{-ik}}{2it_1t_2\sin k}
 =-\frac{1}{2t_1}-i\frac{t_1+t_2\cos k}{2t_1t_2\sin k}.
\end{equation}
Contracting with the two contact amplitudes gives the real self-energy $-\lambda_A\lambda_B\cos\phi/t_1$ and imaginary part $-(u+g)/2$. The waveguide response retains the full energy dependence of $k$, $v$, and $\theta$. No frequency-independent approximation has been made for the waveguide coupling; the separate local loss channels remain Markovian as defined in the model.

For the SSH contacts, the field composition of Sec.~\ref{app:generalpair} gives the following excitation amplitudes even when the qubits have unequal contact strengths:
\begin{equation}
 q_{1L}=\frac{b_{1R}+b_{1L}P r_{2L}t_{1L}/Z}{D_1},\qquad
 q_{2L}=\frac{b_{2R}p t_{1L}}{D_2Z}.
 \label{eq:qapp}
\end{equation}

\section{Contact optimization at fixed directional transmission}
\label{app:optimality}
\label{sec:optimality}
The common phase adjustment gives a candidate optimum. We now allow each qubit to have its own phase and unequal positive A/B strengths, while keeping the full ratio $t_R(E)/t_L(E)$ fixed. The question is whether these extra contact variations can lower the leading pulse error.

\subsection{Allowed contact variations}
\label{app:contactcoordinates}
For arbitrary fixed SSH contacts, the exact ratio is
\begin{equation}
 \frac{t_R}{t_L}=\prod_{j=1}^2
 \frac{x-d_j+i(\gamma-G_j)/2}{x-d_j+i(\gamma+G_j)/2},
 \label{eq:generalratio}
\end{equation}
where $d_j=\omega_j-E_0-\lambda_{A,j}\lambda_{B,j}\cos\phi_j/t_1$ is the shifted transition frequency relative to $E_0$. Fixing a function of frequency restricts more parameters than fixing its value at the carrier. Here $G_j$ and $d_j$ are independent of frequency, so equality on an open band interval implies equality of rational functions. A numerator zero can cancel a denominator pole only if their real parts coincide and $\gamma-G_j=\gamma+G_k$, or $G_j=-G_k$. Positive rate differences exclude such cancellation. Matching the poles to Eq.~(\ref{eq:ratio_nominal}), including their multiplicity at $\Delta=0$, fixes $G_1=G_2=G$ and $\{d_1,d_2\}=\{\Delta/2,-\Delta/2\}$.

Apart from interchange of the qubits, the remaining local coordinates are the phase shifts $\nu_j$ and the logarithmic strength imbalances $r_j$, defined by $\lambda_{A,j}/\lambda_{B,j}=e^{2r_j}$. We write
\begin{align}
 \beta_j&=\theta_*-\nu_j,\quad \phi_j=\pi+\beta_j,\quad
 p_j=\frac{Gt_1}{2\sin\beta_j},\nonumber\\
 \lambda_{A,j}&=\sqrt{p_j}e^{r_j},\quad
 \lambda_{B,j}=\sqrt{p_j}e^{-r_j},\nonumber\\
 \omega_j&=E_0+\frac{p_j\cos\phi_j}{t_1}+d_j.
 \label{eq:generalhardware}
\end{align}
The product $p_j$ specifies contact strengths and is distinct from the propagation amplitude $p=e^{-ikL}$. These settings describe calibrated devices. Once chosen, their phases, strengths, and bare frequencies stay fixed during a pulse. The classification keeps the carrier, geometry, and common loss rate fixed and assumes $G_j>0$. Added modes or rate differences of opposite signs are outside this contact family.

The common balanced choice realizes the real-frequency splitting assumed in Eq.~(\ref{eq:realunfold}). To see this directly, set $s(E)=[\theta_*-\theta(E)]/a_\theta$, so $s(E_0+x)=x^2+O(x^3)$. With $q=\nu/a_\theta$, the contact amplitude is
\begin{equation}
 b_R=\frac{i\lambda a_\theta e^{-i(\theta+\beta)/2}}{\sqrt{2v}}
 (s-q)\,\frac{\sin[a_\theta(s-q)/2]}{a_\theta(s-q)/2}.
 \label{eq:sshrealzeros}
\end{equation}
The sine ratio has its analytic value 1 at $s=q$. The remaining prefactors are analytic and nonzero near the carrier, so the two zeros remain real for small positive $q$.

Near the reference contacts, a phase change adds a real constant to the factor $x^2$ in the coupling. A strength imbalance adds an imaginary constant. Expanding Eq.~(\ref{eq:b}) gives
\begin{align}
 c_j&=K_c(x^2-Q_j)+\text{higher orders},\nonumber\\
 K_c&=\frac{Ga_\theta e^{-i\theta_*}}{2\sin\theta_*},\qquad
 Q_j=\frac{\nu_j-2ir_j}{a_\theta},\qquad |K_c|^2=G^2C_4.
 \label{eq:contactnormal}
\end{align}
The complex coordinate $Q_j$ combines these two changes. Its average $\overline Q=(Q_1+Q_2)/2$ describes their shared part, while $Q_1-Q_2$ describes an unequal change. These coordinates are distinct from the scattering denominators $Q$ and $\mathcal Q$. For unequal contacts, the exact left reflection is
\begin{align}
 r_L&=-i\frac{c_1D_2+Pc_2(D_1-i\Sigma_1)}{\mathcal D},\nonumber\\
 \Sigma_j&=G+2u_j,\quad D_j=x-d_j+i(\gamma+\Sigma_j)/2,\nonumber\\
 \mathcal D&=D_1D_2+Pc_1^*c_2 .
 \label{eq:generalreflection}
\end{align}
Equation~(\ref{eq:cascade}) gives this expression for equal as well as unequal contacts. Define $H=G^2+\Delta^2$, $K=2+G\tau-i\tau\Delta$, and $A=(\Delta+iG)/2$. The carrier denominator tends to $-H/4$ as the contacts approach the unsplit lossless setting.

\subsection{Contact changes compatible with sixth-order pulse error}
\label{app:contactscales}
Set $\mathbf Q=(Q_1,Q_2)$ and expand the exact reflected amplitude as $r_L(x)=r_0+r_1x+r_2x^2+\cdots$. Here $r_0$ is the carrier reflected field and $r_1$ is its frequency slope. Each coefficient depends analytically on the four real contact coordinates in $\mathbf Q$ and on $\gamma$. At $\mathbf Q=0$, $r_0=r_1=0$ for sufficiently small $\gamma$, while $r_2=O(\|\mathbf Q\|+\gamma)$ and higher coefficients are bounded. Apart from a common nonzero prefactor, the first two coefficients respond as
\begin{align}
 r_0&=-A(Q_1-Q_2)+O(\|\mathbf Q\|^2+\gamma\|\mathbf Q\|),\nonumber\\
 r_1&=-K\overline Q+B(Q_1-Q_2)
       +O(\|\mathbf Q\|^2+\gamma\|\mathbf Q\|),
 \label{eq:inversemap}
\end{align}
where $B$ is a finite coefficient in this local map. Since $A\ne0$ and $\operatorname{Re}K>0$, the four real components of $r_0,r_1$ determine the four contact coordinates locally. The inverse map is bounded in a sufficiently small neighborhood. Suppose a sequence of nearby devices without gain has $\langle|r_L|^2\rangle=O(\sigma^6)$ with $\gamma=O(\sigma^2)$. By Cauchy--Schwarz, its overlaps with the constant and normalized linear spectral functions must each be $O(\sigma^3)$. These overlaps are
\begin{align}
 r_0+\sigma^2r_2+3\sigma^4r_4+\cdots&=O(\sigma^3),\nonumber\\
 \sigma r_1+3\sigma^3r_3+\cdots&=O(\sigma^3).
\end{align}
The second line gives $r_1=O(\sigma^2)$, and the first gives $r_0=O(\sigma^3)+O(\sigma^2\|\mathbf Q\|)$. Applying the bounded inverse yields $\|\mathbf Q\|\le C[\sigma^2+\sigma^2\|\mathbf Q\|]$, hence $\|\mathbf Q\|=O(\sigma^2)$. Substituting back into the first line of Eq.~(\ref{eq:inversemap}) gives $Q_1-Q_2=O(\sigma^3)$. The quadratic parameter corrections are $O(\sigma^4)$ and do not alter this restriction on the constant reflected field. Thus a nearby device with sixth-order pulse error must satisfy $\overline Q=O(\sigma^2)$ and $Q_1-Q_2=O(\sigma^3)$.

The analytic bounds hold in one fixed neighborhood of the carrier and contacts. Gaussian tails outside it are exponentially small, with their probabilities bounded by flux conservation. The estimates therefore apply uniformly to the nearby device sequences considered here. They do not compare disconnected designs far from the stationary contacts.

\subsection{Adding reflection and independent loss probabilities}
With the scales just obtained, insert Eq.~(\ref{eq:contactnormal}) and $P=1+2i\tau x+O(x^2)$ into Eq.~(\ref{eq:generalreflection}). The denominator is $\mathcal D=-H/4+o(1)$, and
\begin{equation}
 r_L=\frac{4iK_c}{H}\left[Kx(x^2-\overline Q)-A(Q_1-Q_2)\right]+o(\sigma^3),
 \label{eq:normalreflection}
\end{equation}
The common-contact contribution is odd in detuning, while the leading differential contribution is constant. To separate these spectral shapes, set $y=x/\sigma$. Under a unit-variance Gaussian, the Hermite polynomials $H_0=1$, $H_1=y$, $H_2=y^2-1$, and $H_3=y^3-3y$ are orthogonal, with squared norms $1,1,2,6$. Use the scaled coordinates $Z_{\rm c}$ and $d_{\rm c}$ defined in Eq.~(\ref{eq:scaledcontactcoordinates}). Dividing by $\sqrt{C_6}\sigma^3$ and removing its constant phase gives the field in Eq.~(\ref{eq:hermite_field}).
A common contact change adjusts the coefficient of $H_1$. A difference between contacts adds $H_0$. Neither changes the coefficient of $H_3$. The leading coefficient of the integrated reflection probability, after division by $C_6\sigma^6$, is therefore $6+|3-Z_{\rm c}|^2+|d_{\rm c}|^2$.

At leading order the contacts are common in the two loss amplitudes. Using Eq.~(\ref{eq:qapp}) with zero-loss prefactors gives
\begin{equation}
 \gamma\sum_j|q_{jL}|^2
 =\frac{8\gamma GC_4}{H}|x^2-\overline Q|^2+o(\sigma^6).
\end{equation}
Both loss amplitudes have the spectral shape $y^2-Z_{\rm c}=H_2+(1-Z_{\rm c})H_0$. After their squared norms are added, the leading absorption probability divided by $C_6\sigma^6$ is $\rho[2+|1-Z_{\rm c}|^2]$. Emission into either reservoir and reflection into the waveguide are different final outcomes. Their probabilities add, so a contact change cannot cancel loss by interference with the reflected field. The left amplitude difference is $t_L-1=O(\sigma^5)$, whose variance is $O(\sigma^{10})$. Adding these contributions through Eq.~(\ref{eq:decomp}) gives Eq.~(\ref{eq:allknobs}).
The completed square in Eq.~(\ref{eq:allknobs}) isolates the cost of departing from the optimum. An imaginary part of $Z_{\rm c}$ increases $|Z_{\rm c}-z_*|^2$, and unequal contacts add $|d_{\rm c}|^2$. The minimum is attained by real $Z_{\rm c}=z_*$ and $d_{\rm c}=0$: both qubits use the same phase shift and equal A/B strengths. This fixes the leading narrow-pulse optimum within the stated SSH contact geometry and fixed-ratio family.

Higher-order adjustments can still change the finite-width error. For identical contacts and real $q$, analyticity and vanishing odd Gaussian moments give the $O(\sigma^8)$ remainder in Eq.~(\ref{eq:unfolding}). A generic unequal-contact family can contain an $O(\sigma^7)$ correction. Completing squares fixes the leading scaled coordinates while leaving subleading device parameters undetermined.

\subsection{Reflection and absorption at fixed nonzero loss}
\label{app:fixedloss}
At fixed nonzero $\gamma$, reflection and the two independent losses all contribute at fourth order in pulse width. Put $Q_j=\sigma^2Z_j$, $w_j=x^2-Q_j$, and $\mathbf w=(w_1,w_2)^{\mathsf T}$. Define $D_1^0=-\Delta/2+i(G+\gamma)/2$ and $D_2^0=\Delta/2+i(G+\gamma)/2$. The leading amplitudes of the three output channels are linear functions of $\mathbf w$. Their probabilities add to $\mathbf w^\dagger M\mathbf w$, where
\begin{align}
 M&=\frac{G^2C_4}{|D_1^0D_2^0|^2}
 \left[v_R^\dagger v_R+\frac{\gamma}{G}(v_1^\dagger v_1+v_2^\dagger v_2)\right],\nonumber\\
 v_R&=(D_2^0,D_1^0-iG),\quad v_1=(D_2^0,-iG),\quad v_2=(0,D_1^0).
 \label{eq:lossmatrixrows}
\end{align}
The rows represent the reflected field and the two independent loss fields. Adding their squared norms retains each nonnegative contribution.

For an explicit simplification, write $\delta_0=\Delta/2$, $b_0=(G+\gamma)/2$, and $h=\delta_0^2+b_0^2$. The matrix inside brackets has diagonal entries $(1+\gamma/G)h$. Its off-diagonal entry is $(D_2^0)^*[D_1^0-i(G+\gamma)]=(\delta_0-ib_0)(-\delta_0-ib_0)=-h$. Since $|D_1^0D_2^0|^2=h^2$, one obtains
\begin{equation}
 M=\frac{4GC_4}{\Delta^2+(G+\gamma)^2}
 \begin{pmatrix}G+\gamma&-G\\-G&G+\gamma\end{pmatrix}.
 \label{eq:lossmatrix}
\end{equation}
Both eigenvalues are positive for $\gamma>0$. Resolving the common and differential coordinates and averaging the Gaussian yields
\begin{align}
 \frac{\mathcal E_L(0)}{\sigma^4}={}&B_\gamma[2+|\overline Z-1|^2]\nonumber\\
 &+\frac{2GC_4(2G+\gamma)}{\Delta^2+(G+\gamma)^2}|Z_1-Z_2|^2+o(1).
 \label{eq:fixedloss_all}
\end{align}
Here $\overline Z=(Z_1+Z_2)/2$, and the transmitted variance contributes only at higher order. For $\gamma>0$, both the common pattern $w_1=w_2$ and the differential pattern $w_1=-w_2$ carry a positive error cost. Gaussian averaging therefore gives the unique leading minimum $Z_1=Z_2=1$. The same positivity forces $Q_j=O(\sigma^2)$ for any nearby family with fourth-order error. As $\gamma\to0$, the common-pattern cost in this fourth-order expression vanishes. Its first nonzero contribution must then be obtained from the sixth-order expansion.

\subsection{Opposite incidence}
Every allowed device has $t_R=f t_L$ with the same function $f(E)$. In the joint narrow-pulse limit, this common factor leaves the leading dependence on contact adjustments unchanged for right incidence at the carrier delay or its local correction. The right-incidence error also contains the phase distortion and absorption associated with $f$, which can be much larger. At fixed positive loss, the leading contact-dependent term is multiplied by $|f(0)|^2$ when $f(0)\ne0$. The case $\gamma=G$ and $\Delta=0$ has $f(0)=0$ and requires a higher-order expansion. Thus the same contacts can minimize the leading adjustable contribution in both directions without making the total errors equal.

\section{Finite-pulse response and comparisons}
\label{app:finitepulse}
The output delay changes the transmitted phase overlap but cannot restore reflected or absorbed photons.

\subsection{Pulse overlap and the optimum output delay}
\label{app:pulse}
A photon fails to occupy the desired output mode either because it leaves the transmitted channel or because its transmitted pulse is distorted. To separate these contributions, let the target be the freely propagated input delayed by $t$, and write $\widetilde t_\alpha(E;t)=t_\alpha(E)e^{-itx}$. With a normalized spectral weight, $\langle1\rangle=1$, the mode error is
\begin{align}
 1-|\langle\widetilde t_\alpha\rangle|^2
 &=1-\langle|\widetilde t_\alpha|^2\rangle+
   \langle|\widetilde t_\alpha|^2\rangle-|\langle\widetilde t_\alpha\rangle|^2\nonumber\\
 &=\langle R_\alpha+A_\alpha\rangle+
 \langle|\widetilde t_\alpha-\langle\widetilde t_\alpha\rangle|^2\rangle.
\end{align}
Flux conservation identifies the first contribution as reflection plus absorption. The remaining squared norm measures the transmitted component orthogonal to the target mode. The relation holds for either incident direction and any fixed output delay, with no postselection on transmission.

Keep the carrier a fixed distance from the band edges while $\sigma\to0$. The omitted Gaussian tails then give exponentially small corrections, so full-Gaussian moments determine every algebraic order below. At zero loss, the unsplit pair has $\langle R\rangle=15C_6\sigma^6+O(\sigma^8)$, while the left-incidence transmission phase contributes at higher order. For right incidence, changing the output delay by $a$ from its carrier value leaves the leading phase $b_3x^3-a x$. Its mean is zero under the even Gaussian spectrum. Using $\langle x^2\rangle=\sigma^2$, $\langle x^4\rangle=3\sigma^4$, and $\langle x^6\rangle=15\sigma^6$, its leading variance is
\begin{equation}
 \left\langle(b_3x^3-a x)^2\right\rangle
 =15b_3^2\sigma^6-6ab_3\sigma^4+a^2\sigma^2.
\end{equation}
The minimum occurs at $a=3b_3\sigma^2$ and equals $6b_3^2\sigma^6$. The delay removes the part of the cubic phase with the same spectral dependence as $x$. The remaining polynomial $x^3-3\sigma^2x$ has zero overlap with $x$ under the Gaussian weight. At the carrier delay, $a=0$, the contribution is $15b_3^2\sigma^6$. Higher orders shift the local optimum by $O(\sigma^4)$, and the local solution approaches $t_d$ as $\sigma\to0$. At $|\Delta|=\Gamma_0/\sqrt3$, $b_3=0$ while the relative carrier delay is $6/\Gamma_0$.

The pulse integrals for Fig.~\ref{fig:pulse}(a,b) use the Gaussian weight on $|x|\le8\sigma$, within the propagating band, normalized to one. The omitted probability is below $1.3\times10^{-15}$. Figure~\ref{fig:errors} uses the same normalized finite-pulse definition. We integrate reflection, absorption, and the variance in Eq.~(\ref{eq:decomp}) separately to resolve errors much smaller than one. At $\sigma/J=10^{-2}$ and $\Delta=\gamma=0$, the mean reflection probabilities of the equal-$\Gamma_0$ and equal-$\lambda$ references are $6.167\times10^{-5}$ and $2.606\times10^{-5}$, respectively. Appendix~\ref{app:nominalcertificate} gives the nominal split-device comparisons. Appendix~\ref{app:certificate} supplies the remainder and tail bounds used for the continuous contact-error region.

\subsection{Finite-band comparisons of split and unsplit zeros}
\label{app:nominalcertificate}
\subsubsection{Bounds for the nominal devices}
The nominal comparison uses $G=0.36J$, $\Delta=0$, a prescribed pulse width $\sigma=10^{-2}J$, and $\gamma=0$ or $3.6\times10^{-4}J$. The split device is set to $\nu=3a_\theta\sigma^2$, with strengths and bare frequencies from Eq.~(\ref{eq:family}); the unsplit device has $\nu=0$. The same pair of contact settings is used at both loss rates and is fixed during each pulse. Contact errors are treated separately below and in Appendix~\ref{app:contacterrors}.

To compare the devices without optimizing the reference delay, bound the split-device error from above and the unsplit-device error from below. For the split device at zero delay, Eq.~(\ref{eq:decomp}) and $\operatorname{Var}(t_L)\le\langle|t_L-1|^2\rangle$ give $\mathcal E_{L,s}(0)\le\langle U_s\rangle$, where $U_s=R_s+A_s+|t_{L,s}-1|^2$. For the unsplit device, reflection and absorption give $\mathcal E_{L,0}(t)\ge\langle L_0\rangle$ at every delay, with $L_0=R_0+A_0$. The outward-rounded bounds are
\begin{center}
\begin{tabular}{ccc}
 $\gamma/J$ & Split upper bound & Unsplit lower bound\\
 $0$ & $1.7701\times10^{-8}$ & $3.7827\times10^{-8}$\\
 $3.6\times10^{-4}$ & $1.8301\times10^{-8}$ & $3.8021\times10^{-8}$
\end{tabular}
\end{center}
The displayed bounds imply left-error reductions exceeding $2.01\times10^{-8}$ and $1.97\times10^{-8}$, respectively.

For right incidence, both devices use the fixed delay $t_{\rm ref}=22.1542733447565/J$ in Sec.~\ref{sec:finite}. Let $M_0=\langle t_{R,0}e^{-itx}\rangle$ and $\delta M=\langle(t_{R,s}-t_{R,0})e^{-itx}\rangle$. The difference in mode error is
\begin{equation}
 \mathcal E_{R,0}-\mathcal E_{R,s}
 =2\operatorname{Re}(M_0^*\delta M)+|\delta M|^2.
\end{equation}
Evaluating the overlap difference avoids subtracting errors dominated by the same absorption background. At this common fixed delay, the gain exceeds $2.16\times10^{-8}$ at zero loss and $2.10\times10^{-8}$ at $\gamma=3.6\times10^{-4}J$.

The exact-response integrals are enclosed by interval arithmetic on $y=x/\sigma\in[-9,9]$, using 1024 cells for left incidence and 2048 for right incidence. On every cell, the denominator interval excludes zero.

The omitted full-Gaussian probability is below $2.2844\times10^{-19}$. Flux conservation gives $U_s=2-2\operatorname{Re}t_{L,s}\le4$, which bounds its omitted contribution. The omitted reflection-plus-absorption contribution is nonnegative and therefore does not weaken the lower bound.

For right incidence, the omitted overlap contributions are bounded by one tail probability for $M_0$ and twice that probability for $\delta M$. The in-band normalization lies between the retained core probability and one. Including these factors gives bounds for the complete normalized in-band pulse.

\subsubsection{Selected matched-error comparisons}
The sampled comparison uses $t_1=J$, $t_2=J/2$, $L=3$, $G=0.36J$, $\gamma=3.6\times10^{-4}J$, and $\sigma=\sigma_d=10^{-2}J$. For each $\Delta$, the unsplit device has $\nu=0$ and the split device has $\nu=3a_\theta\sigma_d^2$, with nominal strengths and bare frequencies from Eq.~(\ref{eq:family}). In each comparison, both devices receive the same differential phase error $\delta$, strength-imbalance error $\epsilon$, and symmetric frequency splitting $\Delta$. Each error is measured from that device's own nominal contacts, and the bare frequencies remain at their nominal values after applying it. The 27 tested settings are the Cartesian product
\begin{align}
 \delta&\in\{-10^{-3},0,10^{-3}\},\nonumber\\
 \epsilon&\in\{-5\times10^{-4},0,5\times10^{-4}\},\nonumber\\
 \Delta/J&\in\{-9\times10^{-2},0,9\times10^{-2}\}.
 \label{eq:sample_grid}
\end{align}
Both devices use delay $t=0$ for left incidence and the same $t=t_{\rm ref}$ above for right incidence; these delays are held fixed across all samples. The positive decomposition in Eq.~(\ref{eq:decomp}) is integrated using 2401-point Simpson quadrature on $|x|\le9\sigma$, with the retained Gaussian normalized to one. This interval lies inside the propagating band. The omitted full-Gaussian probability is below $2.2844\times10^{-19}$.

Define the sampled improvement as $\Delta\mathcal E_\alpha=\mathcal E_{\alpha,0}-\mathcal E_{\alpha,s}$, where subscripts $0$ and $s$ label unsplit and split contacts, respectively. The smallest observed improvements are
\begin{align}
 \min_{\rm samples}\Delta\mathcal E_L&\simeq1.20\times10^{-8},\nonumber\\
 \min_{\rm samples}\Delta\mathcal E_R&\simeq9.97\times10^{-9}.
 \label{eq:sample_gain}
\end{align}
At the specified common delays, splitting lowers both mode errors for all 27 tested settings. The minima in Eq.~(\ref{eq:sample_gain}) summarize this finite set; no interpolation between the sampled settings is used to establish a continuous bound.

\section{Contact errors and continuous bounds}
\label{app:contacterrors}
Uncompensated contact errors change both the local interference and the emitter resonance shifts. The local expansion identifies the resulting reflection. The full response then gives bounds throughout the finite parameter region in Eq.~(\ref{eq:box}).

\subsection{Differential contact errors produce carrier reflection}
\label{app:perturbations}
The differential contact errors are defined in Eq.~(\ref{eq:errors}), with $s_1=1$ and $s_2=-1$. They impose opposite phase shifts and opposite A/B strength imbalances on the two qubits. The parameter $\delta$ is measured in radians, and $\epsilon$ is dimensionless.

The bare-frequency midpoint is kept at its nominal value from Eq.~(\ref{eq:omega0}). Changes of the contacts therefore leave uncompensated waveguide-induced frequency shifts. Relative to the nominal setting, these are
\begin{equation}
 \ell_j=\frac{\lambda_{A,j}\lambda_{B,j}\cos\phi_j-\lambda^2\cos\phi_0}{t_1},\qquad
 \xi_j=x-s_j\Delta/2+\ell_j.
 \label{eq:ell}
\end{equation}
The shifts remain in $D_j=\xi_j+i(\gamma+u_j+g_j)/2$ because the bare frequencies are not readjusted. We use the same dimensionless phase and imbalance errors, and the same absolute loss rate, for every device. Equal $\epsilon$ gives different dimensional contact changes when the nominal $\lambda$ values differ. The equal-rate reference matches the nominal carrier rate.

At $\Delta=\gamma=0$, the unwanted carrier amplitude is linear in the local errors:
\begin{equation}
 b_{jR}=\frac{\lambda e^{-i\theta_0}}{\sqrt{2v_0}}\left(-2e_j+i\delta_j\right)+O(\delta_j^2,e_j\delta_j),
 \label{eq:berror}
\end{equation}
where $\delta_j=s_j\delta/2$. The leading reflected fields enter with opposite nominal signs and give Eq.~(\ref{eq:flooramp}). Section~\ref{app:errors} derives this carrier amplitude and its positive-loss counterpart.

The ideal sixth-order Gaussian coefficient survives only when the contact-induced carrier probability is much smaller than $15C_6\sigma^6$. With $s=\sin\theta_0$, this requires $\delta^2+4\epsilon^2\ll15s^2C_6\sigma^6$. Dimensionless errors $o((\sigma/J)^3)$ preserve the coefficient; errors of order $(\sigma/J)^3$ preserve the sixth power but change its coefficient.

A comparison with the fourth-order reference allows larger errors, of order $(\sigma/J)^2$. At that scale, carrier reflection competes with the reference's intrinsic fourth-order pulse error. A strength imbalance and the reference's frequency-dependent reflected amplitude lie along the same axis in the complex-amplitude plane, so their signs can make them partly cancel. A phase error contributes along the perpendicular axis and does not produce that cancellation. At $\Delta=\gamma=0$, define $B_c=6/J+2/\Gamma_{0,c}$. Separate phase-only and amplitude-only comparisons give
\begin{align}
 |\delta|&<4B_c\sigma^2/J,\nonumber\\
 \frac{-2-2\sqrt{10}}{3}\frac{B_c\sigma^2}{J}<\epsilon&<
 \frac{-2+2\sqrt{10}}{3}\frac{B_c\sigma^2}{J}.
 \label{eq:tolerance}
\end{align}
The phase bound assumes $\epsilon=0$, while the amplitude interval assumes $\delta=0$. The factor $J$ restores units to the dimensionless $t_1/J=1$ expressions. These leading asymptotic inequalities do not bound a full finite-band interval. Their asymmetry reflects the orientation of the imbalance: positive $\epsilon$ partly suppresses the reference reflection and narrows that side of the advantage window.

At $\sigma=10^{-2}J$ and zero loss, numerical integration places the local equal-$\lambda$ crossings of the adjusted left error at $|\delta|\simeq2.975\times10^{-3}$. The corresponding imbalance crossings are $\epsilon\simeq-2.067\times10^{-3}$ and $1.071\times10^{-3}$. With fixed $\delta=10^{-3}$ and $\epsilon=0$, the advantage disappears below a local width crossing at $\sigma\simeq5.810\times10^{-3}J$. For $\epsilon=5\times10^{-4}$ and $\delta=0$, the crossing is $\sigma\simeq6.843\times10^{-3}J$. Each fixed device uses its locally adjusted output delay, as in Fig.~\ref{fig:errors}.

\subsection{Reflection remaining at the carrier}
\label{app:errors}
The carrier-reflection coefficients below apply at $\Delta=0$. For simultaneous errors and finite splitting, Appendix~\ref{sec:robustness} retains the full response. Let $s=\sin\theta_0$ and $\delta_j=\phi_j-\phi_0$. At the carrier,
\begin{align}
 b_{jR}&=\frac{\lambda e^{-i\theta_0}}{\sqrt{2v_0}}(-2e_j+i\delta_j)+O(\delta_j^2,e_j\delta_j),\nonumber\\
 b_{jL}&=\frac{2i\lambda s}{\sqrt{2v_0}}+O(\delta_j,e_j),\qquad
 \Gamma_0=\frac{2\lambda^2s^2}{v_0}.
\end{align}
At zero loss the local response denominator is $D_j=i\Gamma_0/2+O(\delta_j,e_j)$. Dividing the first-order coupling product by this denominator gives the reflection from qubit $j$:
\begin{equation}
 r_{jL}=-\frac{e^{-i\theta_0}}s(\delta_j+2ie_j).
\end{equation}
At nominal contacts, a left-moving reflected wave crossing the other qubit acquires $t_{jR}=-1$, whereas $t_{jL}=1$ and $P=1$. The two leading contributions therefore enter as $r_L=r_{1L}-r_{2L}+O((|\delta|+|\epsilon|)^2)$, giving Eq.~(\ref{eq:flooramp}). Contact-induced frequency shifts change the denominators at first order, but multiply a reflected numerator already of first order. They affect carrier reflection only at higher order and remain included in the exact response.

For positive loss put $h=\Gamma_0+\gamma$. Each first-order local reflection is reduced by $\Gamma_0/h$, while the second reflected field is weighted by the nominal opposite-direction transmission $(\gamma-\Gamma_0)/h$. Combining the opposite error signs gives a total factor $\Gamma_0^2/h^2$ relative to the zero-loss amplitude. Thus
\begin{align}
 R_L(0)&=\frac{\Gamma_0^4}{h^4}\frac{\delta^2+4\epsilon^2}{s^2}+O((|\delta|+|\epsilon|)^3),\nonumber\\
 1-T_L(0)&=\frac{\Gamma_0(2\Gamma_0+\gamma)}{2h^2}
             \frac{\delta^2+4\epsilon^2}{s^2}+O((|\delta|+|\epsilon|)^3).
 \label{eq:lossfloor}
\end{align}
The second relation follows by expanding the transmitted amplitude to second order, or by adding the excitation loss to reflection. Subtracting the two expressions gives $A_L(0)$. At first order in $\gamma/\Gamma_0$, the absorption coefficient multiplying $(\delta^2+4\epsilon^2)/s^2$ is $5\gamma/(2\Gamma_0)$.

In this comparison, the subscript $s$ labels the unsplit stationary device, and $c$ labels a bond-swapped reference. For the comparison scales, set $\Delta=\gamma=0$ and use dimensionless energies in units of $J$ in the remainder of this subsection. Put $B_s=6+2/\Gamma_0$, $B_c=6+2/\Gamma_{0,c}$, and $a=1/(2t_2E_*)$. The leading amplitudes near the carrier are
\begin{align}
 r_s&=\frac{e^{-i\theta_*}}s[-\delta-2i\epsilon+2iaB_sx^3]+\cdots,\nonumber\\
 r_c&=e^{-i\theta_{0,c}}[-\delta-2i\epsilon+4iB_cx^2]+\cdots.
 \label{eq:quadratures}
\end{align}
For the stationary design $s=1/2$. A phase error is orthogonal to the intrinsic reflected amplitude at this order, whereas an amplitude imbalance has the same quadrature. In the joint limit $\delta=O(\sigma^2)$ with $\epsilon=0$,
\begin{equation}
 \langle R_s\rangle=4\delta^2+o(\sigma^4),\quad
 \langle R_c\rangle=\delta^2+48B_c^2\sigma^4+o(\sigma^4).
\end{equation}
For $\epsilon=O(\sigma^2)$ with $\delta=0$,
\begin{align}
 \langle R_s\rangle&=16\epsilon^2+o(\sigma^4),\nonumber\\
 \langle R_c\rangle&=4\epsilon^2-16B_c\epsilon\sigma^2+48B_c^2\sigma^4+o(\sigma^4).
\end{align}
Solving the leading inequalities and restoring units gives Eq.~(\ref{eq:tolerance}). The linear cross term sets the asymmetric amplitude window. Common contact errors, arbitrary independent errors, and fixed positive loss in the strict $\sigma\to0$ limit require separate comparisons.

\subsection{Bounds on pulse error in the presence of contact errors}
\label{app:certificate}
We bound the pulse error while allowing phase error, strength imbalance, and symmetric qubit detuning to vary simultaneously. All energies in this subsection are measured in units of $J$. The unsplit device requires four amplitudes: reflection $r_L$, the change $h_L=t_L-1$ in transmission, and the emitter excitations $q_{1L},q_{2L}$. For either reference, reflection alone gives a lower bound on the mode error. With $Q=D_1D_2+Pc_1^*c_2$, Eqs.~(\ref{eq:cascade}) and (\ref{eq:qapp}) give
\begin{align}
 r_L&=-i\frac{c_1D_2+Pc_2[D_1-i(u_1+g_1)]}{Q},\nonumber\\
 h_L\equiv t_L-1&=\frac{-i(u_1D_2+u_2D_1)-u_1u_2-Pc_1^*c_2}{Q},\nonumber\\
 q_{1L}&=\frac{b_{1R}D_2-ib_{1L}Pc_2}{Q},\qquad
 q_{2L}=\frac{b_{2R}p(D_1-iu_1)}{Q}.
 \label{eq:encloseresponse}
\end{align}

Interval bounds can become unnecessarily wide if two occurrences of the same variable are treated as independent. We combine the amplitudes algebraically before bounding them to retain the cancellations responsible for small reflection. For example, with $q_\theta=e^{-i\theta}$,
\begin{equation}
 c_1-c_2=\frac{\lambda^2}{v}\left[\epsilon(1-q_\theta^2)-2(1-\epsilon^2/4)\sin\phi_0\sin(\delta/2)q_\theta\right].
\end{equation}
The reflection numerator can be evaluated as $(c_1-c_2)D_2+c_2M$, where
\begin{align}
 M={}&2x+\ell_1+\ell_2+i\gamma
 +\frac{i}{2}\left[(u_2+g_2)-(u_1+g_1)\right]\nonumber\\
 &+(P-1)\left[D_1-i(u_1+g_1)\right].
\end{align}
The identity $D_1D_2=\bar D^2-(D_1-D_2)^2/4$, with $\bar D=(D_1+D_2)/2$, likewise preserves the fixed midpoint. The cancellations are therefore retained before any interval bounds are applied.

\subsubsection{A reflection component orthogonal to constant fields}
Section~\ref{sec:continuous} defines the unnormalized truncated moments $M_n^{(T)}$ and the function $H_{2,T}=y^2-\mu_T$ on $|y|\le T=6$. Its squared norm is $V_T=M_4^{(T)}-[M_2^{(T)}]^2/M_0^{(T)}$. Equation~(\ref{eq:projection}) removes every constant and odd part of the reference reflection. To make this lower bound useful throughout the contact-error region, we retain the changed frequency-dependent coefficients and bound their Taylor remainder.

Each response is enclosed as a degree-15 Taylor polynomial plus a remainder,
\begin{equation}
 f(x)=\sum_{n=0}^{15}f_nx^n+{\cal R}_{16}(x).
 \label{eq:Taylor}
\end{equation}
Real interval Taylor recurrences cover twelve adjacent $\Delta$ intervals of width $1.5\times10^{-2}$ whose union is exactly $[-9\times10^{-2},9\times10^{-2}]$. The full phase, amplitude, and width intervals in Eq.~(\ref{eq:box}) are retained on each one. On each splitting subinterval the stationary reflection coefficients are evaluated for $\widetilde r_L=\chi r_L$, where
\begin{equation}
 \chi=e^{i\theta_*}\frac{\Gamma_0+i\Delta_{\rm mid}}
 {\sqrt{\Gamma_0^2+\Delta_{\rm mid}^2}},\qquad \Gamma_0=0.36,
 \label{eq:rephase}
\end{equation}
and $\Delta_{\rm mid}$ is the subinterval midpoint. Multiplying $r_L$ by $\chi$ rotates its constant phase and leaves its magnitude unchanged. Choosing this phase separately for each $\Delta$ subinterval makes the interval estimates tighter. It changes neither the polynomial and remainder norms nor the complex-domain bound. No physical output phase or delay is adjusted, and $h_L$ and $q_{jL}$ are unchanged. The reference reflection keeps its original phase. Its projection lower bound uses
\begin{align}
 &\operatorname{Re}\langle H_{2,T},r_c\rangle_T\nonumber\\
 &\quad\ge\inf\operatorname{Re}\sum_{\substack{2\le n\le14\\n\ {\rm even}}}
 r_{c,n}\sigma^n(M_{n+2}^{(T)}-\mu_T M_n^{(T)})
 -\sqrt{V_T}\,\|{\cal R}_{16}\|.
 \label{eq:projectionpoly}
\end{align}
The right-hand side is positive on every reference subinterval. For the stationary upper bound, squared polynomial norms retain all even-moment cross terms. Minkowski's inequality adds the remainder norm before squaring. Summing the resulting squared norms of $r_L$, $h_L$, and $\sqrt\gamma q_{jL}$ gives the upper bound in Eq.~(\ref{eq:certupper}).

\subsubsection{Analytic bound on the Taylor remainder}
The Taylor remainder is controlled by the response at nearby complex frequencies. We continue the exact response analytically in $x$ using
\begin{align}
 \cos k&=\frac{E^2-t_1^2-t_2^2}{2t_1t_2},\qquad
 \sin k=\sqrt{1-\cos^2k},\nonumber\\
 q_\pm&=\frac{t_1+t_2\cos k\pm it_2\sin k}{E}.
\end{align}
Choose the square root by analytic continuation from $\sin k(E_0)>0$ on the physical band. At real frequency, $q_+=q_-^*$. After continuing $x$ to a complex value, the analytic continuation of the conjugated real-frequency factor is $q_+(x)$; it is not obtained by conjugating $x$. This keeps the response analytic in $x$, as required for the Cauchy remainder bound.

We cover each complex-frequency disk by finitely many rectangles in its real and imaginary parts. Over all allowed contact parameters, these rectangles determine a consistent continuation of the square root and keep every response denominator away from zero. The disk radii are $\rho=0.1$ for the unsplit and equal-$\lambda$ devices and $\rho=8\times10^{-2}$ for the equal-$\Gamma_0$ reference. The amplitude bounds are
\begin{equation}
 |r_L|,|h_L|\le5,\qquad |q_{1L}|,|q_{2L}|\le25.
\end{equation}
These disks serve only to bound the analytic response; their radii do not further restrict the physical contact-error region. The nonzero denominator bounds exclude poles within them. For the applicable amplitude bound $B_f=5$ or 25, Cauchy's inequality gives $|f_n|\le B_f/\rho^n$. Summing the omitted powers for $|x|\le a=6\sigma_{\max}=6.6\times10^{-2}<\rho$ yields
\begin{align}
 |{\cal R}_{16}(x)|&\le\frac{B_f|x|^{16}}{\rho^{16}(1-a/\rho)},\nonumber\\
 \|{\cal R}_{16}(\sigma y)\|&\le
 \frac{B_f\sigma_{\max}^{16}\sqrt{M_{32}^{(T)}}}{\rho^{16}(1-a/\rho)}.
 \label{eq:remainder}
\end{align}
This supplies a uniform remainder norm for Eqs.~(\ref{eq:projectionpoly}) and (\ref{eq:certupper}).

\subsubsection{Gaussian tails and pulse normalization}
The truncated Gaussian moments obey
\begin{equation}
 M_{2n}^{(T)}=(2n-1)M_{2n-2}^{(T)}-2T^{2n-1}\varphi(T),\qquad M_{2n+1}^{(T)}=0.
\end{equation}
Standard Gaussian-tail inequalities bound the retained weight $M_0^{(T)}$ and the omitted probability. At $T=6$ the omitted probability is below $2.02530\times10^{-9}$. Since the retained interval lies entirely in the band, the full in-band normalization obeys $M_0^{(T)}\le Z_{\rm band}\le1$. Flux conservation gives
\begin{equation}
 |r_L|^2+A_L+|t_L-1|^2=2-2\operatorname{Re}t_L\le4.
\end{equation}
For the unsplit upper bound, divide the retained-interval result by $M_0^{(T)}$ and add the tail allowance below $8.102\times10^{-9}$. For the reference lower bound, omitting positive reflection outside the retained interval lowers the integral. Normalization by $Z_{\rm band}\le1$ can only increase it. The comparison therefore applies to the complete normalized in-band pulse.

The interval evaluations use outward-rounded real intervals in \texttt{mpmath.iv}. Together with the disk covers, analytic remainder estimates, and Gaussian-tail bounds above, they yield the uniform error bounds reported in Appendix~\ref{sec:robustness}.

\subsection{Performance over a finite range of contact errors}
\label{sec:robustness}

We compare the unsplit device with both bond-swapped references over the continuous error region in Eq.~(\ref{eq:box}). The differential phase and A/B strength errors are defined in Eq.~(\ref{eq:errors}). The bare frequencies stay at their nominal values, so the contact-induced shifts in Eq.~(\ref{eq:ell}) remain in the response.

The projection in Appendix~\ref{app:certificate} gives a positive reflection lower bound for both references throughout Eq.~(\ref{eq:box}). It retains the frequency-dependent effects of contact errors and the analytic remainder. In this comparison, the subscript $s$ labels the unsplit stationary device, and $c$ labels a bond-swapped reference. For the unsplit device, fix the output delay to zero. Equation~(\ref{eq:decomp}) and $\operatorname{Var}(t_L)\le\langle|t_L-1|^2\rangle$ give
\begin{equation}
 1-F_{L,s}(0)\le\left\langle|r_{L,s}|^2+\gamma\sum_j|q_{jL,s}|^2+|t_{L,s}-1|^2\right\rangle.
 \label{eq:certupper}
\end{equation}
For either reference and any output delay,
\begin{equation}
 1-F_{L,c}(t)\ge\langle R_{L,c}+A_{L,c}\rangle\ge\langle R_{L,c}\rangle.
 \label{eq:certlower}
\end{equation}
A gap between these bounds persists at any reference delay: adjusting the transmitted phase cannot recover a reflected photon.

Including the analytic remainders and Gaussian tails gives the uniform bounds in Eq.~(\ref{eq:certificate}). The unsplit device loses less of the desired output mode than either reference loses through reflection alone.
The left-error reduction exceeds $6.08\times10^{-6}$ relative to the equal-$\Gamma_0$ reference and $2.21\times10^{-6}$ relative to the equal-$\lambda$ reference [Fig.~\ref{fig:joint}(a)]. Each inequality holds throughout Eq.~(\ref{eq:box}). The unsplit device can therefore have different contact errors and a different pulse width from the reference, provided both remain inside this region.

The transmission phases also retain a difference between directions. We characterize it by the carrier group delay $t_\alpha^{\rm gr}=\left.\partial_E\arg t_\alpha(E)\right|_{E=E_0}$, using the same free-propagation reference for both directions. With uncorrected contacts, write $\xi_j=x-d_j$, where $d_j$ includes the uncompensated shift in Eq.~(\ref{eq:ell}). The exact directional ratio is
\begin{equation}
 \frac{t_R}{t_L}=\prod_{j=1}^2\frac{\xi_j+i(\gamma-G_j)/2}{\xi_j+i(\gamma+G_j)/2}.
 \label{eq:ratio_imperfect}
\end{equation}
Set $\zeta_j=\xi_j(0)$. Differentiating Eq.~(\ref{eq:ratio_imperfect}) gives
\begin{align}
 t_R^{\rm gr}-t_L^{\rm gr}=2\sum_{j=1}^2\bigg[&\frac{G_j-\gamma}{4\zeta_j^2+(G_j-\gamma)^2}\nonumber\\
 &+\frac{G_j+\gamma}{4\zeta_j^2+(G_j+\gamma)^2}\bigg].
 \label{eq:delaydiff}
\end{align}
The same free propagation is removed from both directions. Over Eq.~(\ref{eq:box}), the delay difference lies in Eq.~(\ref{eq:responsebound}), while $|t_L(E_0)|>0.9394$ and $|t_R(E_0)|>0.9356$. The nonzero carrier amplitudes keep both phase delays well defined, and their difference remains larger than $10/J$.

The improvement from deliberate zero splitting is about $10^{-8}$ for the chosen pulse, whereas the uncorrected-contact background can reach order $10^{-5}$ in Eq.~(\ref{eq:box}). Resolving the splitting benefit therefore requires coordinated calibration of phases, strengths, and bare frequencies. Appendix~\ref{app:nominalcertificate} finds lower errors for the split device at all 27 matched-error settings, using a common output delay for each incidence direction. The continuous bounds in this subsection concern the unsplit device. They show that a finite-pulse advantage can survive after contact errors destroy the ideal sixth-order law.

\end{document}